\documentclass[12pt]{article}
\usepackage{amssymb}
\usepackage{amsmath}
\usepackage{thophys}
\usepackage{latexsym}
\usepackage{epsfig}
\usepackage{float}

\def\u{\mathbf{u}}
\def\d{\mathbf{d}}
\def\m{\mathbf{m}}
\def\S{\mathbf{S}}
\def\G{\mathbf{G}}
\def\V{\mathbf{V}}

\def\Si{\mbox{\boldmath{$\Sigma$}}}

\def\deZ{\mathbf{\delta Z}}
\def\dem{\mathbf{\delta m}}

\def\tr{\mathrm{Tr}}
\def\s{\mathrm{s}}
\def\c{\mathrm{c}}

\begin{document}

\renewcommand{\thefootnote}{\fnsymbol{footnote}}
\thispagestyle{empty}
\begin{titlepage}
\begin{flushright}
hep-ph/9810350\\
TTP 98--34 \\
\today
\end{flushright}

\vspace{0.3cm}
\boldmath
\begin{center}
\Large\bf  The Renormalization Group Evolution of the CKM Matrix
\end{center}
\unboldmath
\vspace{0.8cm}

\begin{center}
{\large Christopher Balzereit, Thomas Hansmann,\\        
          Thomas Mannel and Benedikt Pl\"umper\footnote[1]{Address after Feb.
	1999: DESY - Theory Group, Notkestra\ss{}e 85, D-22603 Hamburg}}\\
\vspace{.5cm}

{\sl Institut f\"{u}r Theoretische Teilchenphysik,
     Universit\"at Karlsruhe,\\ D -- 76128 Karlsruhe, Germany} 
\end{center}

\vspace{\fill}

\begin{abstract}
\noindent
We compute the renormalization of the complete CKM matrix
in the $\overline{MS}$ scheme and perform a renormalization group analysis 
of the CKM parameters. The calculation is simplified by 
studying only the Higgs sector, which for the $\beta$-function   
of the CKM matrix is at one loop the same as in the full Standard Model. 
The renormalization group flow including QCD corrections 
can be computed analytically 
using the hierarchy of the CKM parameters and the large mass 
differences between the quarks. While the evolution of 
the Cabibbo angle is tiny
$V_{ub}$ and $V_{cb}$ increase sizably.
We compare our results with the ones in the full Standard Model.
\end{abstract}
\end{titlepage}

\renewcommand{\thefootnote}{\arabic{footnote}}
\newpage

\section{Introduction}
In the Standard Model (SM) and in all possible extensions the origin of 
flavor mixing lies in the Higgs sector and thus belongs to its least
understood part. 
While in the quark sector this phenomenon is parametrized by 
the Cabibbo-Kobayashi-Maskawa (CKM) matrix \cite{ckm}, it is still not 
clear whether a similar effect exists for the leptons. Here mixing 
can only happen once the neutrinos have masses, for which recently
some evidence has been given \cite{neutrino}.

As a first step in understanding the origin of the CKM matrix
it is useful to compute its renormalization group evolution, 
since one may hope that some unknown physics fixes  
a CKM matrix or, equivalently, mass matrices for quarks and leptons 
at a high scale $\Lambda$.
Thus its structure can give some hint on the overlying theory which 
produces this CKM matrix as an effective coupling. 
The CKM matrix elements are measured at hadronic scales 
of a few GeV, or maybe at the electroweak scale if they are 
extracted from $W$ decays one day. 

Studies of the renormalization of  the CKM matrix can already
be found in the literature. The one loop contributions 
to the CKM matrix have been computed in \cite{denner-sack} in the on-shell
scheme. It has been found that the corrections are small and hence
they have been ignored in all analysis. However, in \cite{denner-sack}
the renormalization group flow has not been investigated. In a couple of other 
papers \cite{arason,olechowski,lindner1,lindner2,babu}
the renormalization group flow of the parameters of the Higgs sector
has been studied. Since  the system of renormalization group equations is quite
complicated these studies have been performed numerically. 
If one considers non-SM scenarios mixing can also occure
in the leptonic sector which has been studied in \cite{kniehl}.

In the present paper we present a renormalization group study
of the CKM matrix, where we simplify matters in such a way
that we may even construct an analytic solution of the renormalization group
equations which is an excellent approximation below the GUT scale of $10^{15}\,
\mathrm{GeV}$. Thus we restrict ourselves to a one-loop analysis
in the limit of vanishing elektroweak 
gauge couplings.
Consequently for the renormalization of the CKM matrix 
only the Higgs sector remains but, as we shall see,
the full SM  result is reproduced at one loop.

In the next section we shall ``ungauge'' the elektroweak part of the
SM
by taking the limit of vanishing gauge coupling in an appropriate 
way. Since the renormalization group evolution of the quark masses 
is mainly driven by strong interactions, 
the QCD part remains as  in the full SM.
Section 3 discusses the renormalization of this pure
Higgs sector. In particular it is shown that due to a Ward identity 
the renormalization 
of the CKM matrix only involves the wave function renormalization
matrices of the left handed quarks.
This is true to all orders in the loop expansion.
In section 4 we formulate the renormalization group
equations for the CKM parameters and the masses and solve them analytically in
a certain approximation. In section 5 we check the quality of our analytic solution
by comparing with the full SM, i.e. with non-vanishing electroweak
couplings. Finally we discuss our results and conclude.

\section{Higgs Sector and Flavor Mixing}
Flavor mixing is entirely generated by the Higgs sector
and the physics of this effect should be understandable without
the complications of the gauge theory. Thus we choose to ``ungauge''
$SU(2)\otimes U(1)_Y$ in the following way. We take the limit
$g_1 \to 0$ and $g_2 \to 0$ ($g_1$ and $g_2$ being the 
$SU(2)$ and $U(1)_Y$ couplings respectively) keeping the vacuum
expectation value of the Higgs field fixed. Furthermore, the
ratio $g_1/g_2$ defining the weak mixing angle does not enter
our consideration.
In this limit the longitudinal modes of the weak bosons appear
as massless scalar fields, namely as the Goldstone bosons of the
spontaneously broken $SU(2)_L$, while the transverse degrees of freedom 
decouple. 

We shall group all known quarks and leptons into
left and right handed doublets according to 
\begin{eqnarray}
\begin{array}{ccc}
L_u = \left( \begin{array}{c} u \\ d \end{array} \right)_L \quad
L_c = \left( \begin{array}{c} c \\ s \end{array} \right)_L \quad
L_t = \left( \begin{array}{c} t \\ b \end{array} \right)_L 
\end{array} 
\end{eqnarray}
\begin{eqnarray}
\begin{array}{ccc}
R_u = \left( \begin{array}{c} u \\ d \end{array} \right)_R \quad
R_c = \left( \begin{array}{c} c \\ s \end{array} \right)_R \quad
R_t = \left( \begin{array}{c} t \\ b \end{array} \right)_R 
\end{array}
\end{eqnarray}
\begin{eqnarray}
\begin{array}{ccc}
L_e = \left( \begin{array}{c} \nu_e \\ e \end{array} \right)_L \quad
L_\mu = \left( \begin{array}{c} \nu_\mu \\ \mu \end{array} \right)_L \quad
L_\tau = \left( \begin{array}{c} \nu_\tau \\ \tau \end{array} \right)_L 
\end{array}
\end{eqnarray}
\begin{eqnarray}
\begin{array}{ccc}
R_e = \left( \begin{array}{c} \nu_e \\ e \end{array} \right)_R \quad
R_\mu = \left( \begin{array}{c} \nu_\mu \\ \mu \end{array} \right)_R \quad
R_\tau = \left( \begin{array}{c} \nu_\tau \\ \tau \end{array} \right)_R 
\end{array} \,.
\end{eqnarray}
Note that we have also introduced right handed neutrino fields
in order to complete the right handed leptonic doublets. We shall
write the Higgs sector of the SM first as a linear sigma model
which has a full $SU(2)_L \otimes SU(2)_R$ symmetry. Upon 
spontaneous breaking this symmetry is reduced to $SU(2)_{L+R}$
corresponding to the custodial symmetry of the SM.
As we shall see, flavor mixing is related to the explicit breaking
of this symmetry and hence we introduce some explicit breaking 
later.

Under this symmetry the left handed leptons and quarks
transform as a $(2,0)$ while the right handed components
are assigned to the $(0,2)$ representation. Furthermore, to make 
contact with the weak hypercharge of the SM
we postulate another $U(1)$ symmetry under which we
assign the following charges 
\begin{equation}
\begin{array}{cccr}
L_q & \stackrel{U(1)}{\longrightarrow} & e^{i\,(1/3)\,\omega} L_q 
& \quad (q = u, c, t) \\
L_l & \stackrel{U(1)}{\longrightarrow} & e^{i\,(-1)\,\omega} L_l
& \quad (l = e, \mu, \tau) \\
R_q & \stackrel{U(1)}{\longrightarrow} & e^{i\,(1/3)\,\omega} R_q
& \quad (q = u, c, t) \\
R_l & \stackrel{U(1)}{\longrightarrow} & e^{i\,(-1)\,\omega} R_l
& \quad (l = e, \mu, \tau)\,. 
\end{array}
\end{equation}
The four Higgs fields
are gathered in a $2 \times 2$ matrix according to
\begin{equation}
\Bbb{H} =  \left(
\begin{array}{cc} \varphi_0 -i\chi & \sqrt{2}\,\phi^+ \\
-\sqrt{2}\,\phi^- & \varphi_0 + i\chi \end{array} \right) 
\end{equation}
transforming in an obvious way under $(2,\bar 2)$
while it is invariant under this additional $U(1)$.
The Higgs fields are governed by the standard lagrangian of the
linear sigma model
\begin{equation}
\mathcal L = \frac{1}{2} \, \mathrm{Tr} \, \left[ \left(\partial_\mu \Bbb{H}^\dagger \right)
\left(\partial^\mu \Bbb{H} \right) \right]
-\frac{\lambda}{64} \left[ \mathrm{Tr} \left( \Bbb{H}^\dagger \, \Bbb{H} \right)
\right]^2
+ \frac{\mu^2}{4} \mathrm{Tr} \left( \Bbb{H}^\dagger \, \Bbb{H} \right)
\end{equation}
which exhibits spontaneous symmetry breaking, if $\mu^2 > 0$.
We choose the vacuum expectation value such that at tree level
\begin{equation}
\varphi_0 = v + H\,,
\qquad v = \sqrt{\frac{4 \mu^2}{\lambda}} \,.
\end{equation}
This choice yields a breaking term proportional to the 
unit matrix which breaks the full $SU(2)_{L}
\otimes SU(2)_{R}$ symmetry down to the diagonal $SU(2)_{L+R}$ 
which is usually called custodial $SU(2)$.  
The field $H$ is the physical Higgs field while the other
fields $\chi$, $\phi^\pm$ are the Goldstone bosons and 
correspond to the longitudinal degrees
of freedom of the $Z_0$ and $W^\pm$.

The only possible renormalizable coupling terms of the 
Higgs fields to the matter fields invariant under
$SU(2)_L \otimes SU(2)_R$ and the additional $U(1)$ are
\begin{equation} \label{ww1}
\mathcal L^{(0)}_{ffH}= - \sum_{A,B=u,c,t} \bar{L}_A \Bbb{H}\, R_B G_{AB} -
\sum_{a,b=e,\mu,\tau} \bar{L}_a \Bbb{H}\, R_b g_{ab} + \mathrm{h.c.}\,.
\end{equation}
Obviously the matrices of Yukawa couplings $G_{AB}$ and $g_{ab}$ can be
diagonalized by the usual biunitary transformation 
without any effects on the other terms in the lagrangian and 
hence no flavor mixing can appear as long as the custodial $SU(2)_{L+R}$ 
remains unbroken.  

Different masses for up- and down-type quarks as well as the coupling to
hypercharge break the custodial $SU(2)$ in the full SM. In the ``ungauged''
model we introduce the breaking of $SU(2)_{L+R}$ by an additional coupling
of the form:
\begin{equation} \label{ww2}
\mathcal L^{(1)}_{ffH} = - \sum_{A,B=u,c,t} \bar{L}_A \Bbb{H}\, \sigma_3 R_B \tilde{G}_{AB} -
\sum_{a,b=e,\mu,\tau} \bar{L}_a \Bbb{H}\, \sigma_3 R_b \tilde{g}_{ab}
+ \mathrm{h.c.} \,\,\, ,
\end{equation}
which also breaks $SU(2)_R$ down to a $U(1)_R$.
In this way the relation between the breaking of custodial $SU(2)$ and mixing
becomes transparent.

The symmetry needed for the elektroweak part of the SM
is still present as a combination of the $U(1)$ introduced above
and this $U(1)_R$.
Hence we introduce a hypercharge
$U(1)_Y$, under which the fermion dubletts transform according to 
\begin{equation}
\begin{array}{cccl}
L_q & \stackrel{U(1)_Y}{\longrightarrow} & e^{i\,(1/3)\,\omega} L_q 
& \quad (q = u, c, t) \\
L_l & \stackrel{U(1)_Y}{\longrightarrow} & e^{i\,(-1)\,\omega} L_l
& \quad (l = e, \mu, \tau) \\
R_q & \stackrel{U(1)_Y}{\longrightarrow} & e^{i\,(1/3 + \sigma_3)\,\omega} R_q
& \quad (q = u, c, t) \\
R_l & \stackrel{U(1)_Y}{\longrightarrow} & e^{i\,(-1 + \sigma_3)\,\omega} R_l
& \quad (l = e, \mu, \tau)\,.
\end{array}
\end{equation}

Due to the explicit breaking of custodial $SU(2)$ the up and down type
quarks aquire different mass matrices defined as
\begin{eqnarray}
G_{u,AC} &\equiv&  v \left( G_{AC} + \tilde{G}_{AC} \right) \nonumber\\
G_{d,BD} &\equiv&  v \left( G_{BD} - \tilde{G}_{BD} \right).
\end{eqnarray}
In the following we shall discuss only quarks for which we introduce 
the compact notation
\begin{equation}
\begin{array}{cc}
\u_{L,R} = \left(\begin{array}{c} u \\ c \\ t \end{array} \right)_{L,R}
&
\d_{L,R} = \left(\begin{array}{c} d \\ s \\ b \end{array} \right)_{L,R} .
\end{array}
\end{equation}
The mass terms for the quarks take the form
\begin{equation}
{} - \bar{\u}_L \G_u \u_R - \bar{\u}_R \G_u^\dagger \u_L -
\bar{\d}_L \G_d \d_R - \bar{\d}_R \G_d^\dagger \d_L
\end{equation}
which upon diagonalization yields the mass spectrum of the quarks
\begin{eqnarray} \label{biunitaer}
\m_u &=& \S_u^{L\dagger} \,\G_u \,\S_u^R = \mathbf{diag}\,
\left(m_u, m_c, m_t \right) \nonumber\\
\m_d &=& \S_d^{L\dagger}\, \G_d \,\S_d^R = \mathbf{diag}\,
\left(m_d, m_s, m_b \right) 
\end{eqnarray}
where $m_{i}> 0$.
As in the full SM the CKM matrix is given by
\begin{equation}
\V \equiv \S_u^{L\,\dagger} \, \S_d^L .
\end{equation}
In this basis of mass eigenstates we find in the broken phase
\begin{eqnarray} \label{lagrangedichte}
\mathcal{L} &=&
\mathcal{L}_{kin}^{Higgs} + \mathcal{L}_{kin}^{quarks}
+ \mathcal{L}_{int}^{Higgs} + \mathcal{L}_{int}^{neutral} +
\mathcal{L}_{int}^{charged} \\
\nonumber \\
\mathcal{L}_{kin}^{Higgs} &=&
\frac{1}{2} H \left( \Box - M_H^2 \right) H +
\frac{1}{2} \chi \left( \Box - M_\chi^2 \right) \chi +
\phi^+ \left( \Box - M_\chi^2 \right) \phi^-  \\
\mathcal{L}_{kin}^{Quarks} &=&
\bar{\u}_L i \fmslash{\partial} \u_L + \bar{\u}_R i \fmslash{\partial}\u_R
- \bar{\u}_L \m_u \u_R - \bar{\u}_R \m_u \u_L \nonumber \\
&& + \bar{\d}_L i \fmslash{\partial} \d_L + \bar{\d}_R i \fmslash{\partial}\d_R
- \bar{\d}_L \m_d \d_R - \bar{\d}_R \m_d \d_L \\
\mathcal{L}_{int}^{Higgs} &=& - v  M_\chi^2 H - \frac{M_H^2 - M_\chi^2}{2v}
\bigg[ H^3 + H\chi^2 + 2 H \phi^+\phi^- \bigg] \nonumber \\
&& - \frac{M_H^2 - M_\chi^2}{8v^2} \bigg[ H^4 + \chi^4 +
4\left(\phi^+\phi^-\right)^2 + 2 H^2\chi^2\nonumber \\
&& + 4H^2\phi^+\phi^- +
4\chi^2\phi^+\phi^- \bigg]  \\
\mathcal{L}_{int}^{neutral} &=& 
- \frac{1}{v} \bar{\u}_L \m_u \u_R \left(H-i\chi\right)
- \frac{1}{v} \bar{\u}_R \m_u \u_L \left(H+i\chi\right) \nonumber \\
&& - \frac{1}{v} \bar{\d}_L \m_d \,\d_R \left(H-i\chi\right)
- \frac{1}{v} \bar{\d}_R \m_d \,\d_L \left(H+i\chi\right) 
\label{L-int-neutral}\\
\mathcal{L}_{int}^{charged} &=& 
- \frac{\sqrt{2}}{v} \bar{\u}_L \V \m_d \,\d_R \phi^+
+ \frac{\sqrt{2}}{v} \bar{\u}_R \m_u \V \d_L \phi^+ \nonumber \\
&& + \frac{\sqrt{2}}{v} \bar{\d}_L \V^\dagger \m_u \u_R \phi^-
- \frac{\sqrt{2}}{v} \bar{\d}_R \m_d \V^\dagger \u_L \phi^- .
\label{L-int-geladen}
\end{eqnarray}
For the purpose of renormalization we have also introduced
a mass for the Goldstone bosons such that the masses of the Higgs particles are
\begin{eqnarray}
M_H^2 &\equiv& \frac{3}{4} \lambda v^2 - \mu^2 \nonumber \\
M_\chi^2 &\equiv& \frac{1}{4} \lambda v^2 - \mu^2.
\end{eqnarray}
At tree level $M^{2}_{\chi}=0$ but for renormalization it is advantageous
to keep $v$ as an independent parameter. Equation 
(\ref{L-int-neutral}) represents 
the neutral currents
and the interactions with the physical Higgs. These contributions - as in the
full SM - do not induce quark mixing. The charge current 
interactions (\ref{L-int-geladen}) 
involve the CKM matrix $\V$ and are the source of flavor mixing.

\section{Renormalization}

The Higgs sector is introduced in such a way that 
the full as well as the ``ungauged'' SM is renormalizable.
We are aiming here at the one loop renormalization of the CKM matrix which
can be obtained from the quark self energies only.
This is due to a Ward identity which actually holds to all orders and even
in the full SM. It is a consequence of $SU(2)_L$ and 
is derived in the unbroken phase. In this phase global $SU(2)_{L}$ is a 
manifest symmetry which translates into the Ward identities
\begin{eqnarray}
\frac{\delta \Gamma}{\delta \phi^{-}_{0}}\, \varphi_{0,0}
 + \sqrt{2}\,\biggl[
\frac{\delta \Gamma}{\delta {\bf d}_{L,0}} {\bf u}_{L,0}
+{\bf \bar d}_{L,0}\frac{\delta \Gamma}{\delta {\bf \bar u}_{L,0}}
\biggr] &=& 0 \nonumber \\
\frac{\delta \Gamma}{\delta \phi^{+}_{0}}\, \varphi_{0,0}
 - \sqrt{2}\,\biggl[
\frac{\delta \Gamma}{\delta {\bf u}_{L,0}} {\bf d}_{L,0}
+{\bf \bar u}_{L,0}\frac{\delta \Gamma}{\delta {\bf \bar d}_{L,0}}
\biggr] &=& 0 \label{wardunbroken}
\end{eqnarray}
for the generating functional $\Gamma$ of one particle irreducible
Greensfunctions. The functions $\phi^{\pm}_{0}$ and ${\bf u/d}_{L,0}$
have to be regarded as sources for the corresponding fields.
(\ref{wardunbroken}) holds for the bare\footnote{Bare parameters and 
fields will be labeled with an additional 
subscript 0.}
Higgs field
and bare elektroweak eigenstates 
and we have suppressed terms which will not appear
as external states. In the broken phase the same Ward identities hold
with $\varphi_{0,0}$ replaced by $v_{0} + H_{0}$
\begin{eqnarray}
\frac{\delta \Gamma}{\delta \phi^{-}_{0}}\, v_{0}
 + \sqrt{2}\,\biggl[
\frac{\delta \Gamma}{\delta {\bf d}_{L,0}} {\bf u}_{L,0}
+{\bf \bar d}_{L,0}\frac{\delta \Gamma}{\delta {\bf \bar u}_{L,0}}
\biggr] &=& 0 \nonumber \\
\frac{\delta \Gamma}{\delta \phi^{+}_{0}}\, v_{0}
 - \sqrt{2}\,\biggl[
\frac{\delta \Gamma}{\delta {\bf u}_{L,0}} {\bf d}_{L,0}
+{\bf \bar u}_{L,0}\frac{\delta \Gamma}{\delta {\bf \bar d}_{L,0}}
\biggr] &=& 0 \,.\label{wardbroken}
\end{eqnarray}
We have ommited the term with the physical Higgs field since it 
does not contribute to the renormalization of the charged current.
Transforming to bare mass eigenstates according to
\begin{eqnarray} 
\u_{L/R,0} &\to& \S_{u,0}^{L/R} \,\u_{L/R,0} \nonumber\\ 
\d_{L/R,0} &\to& \S_{d,0}^{L/R} \,\d_{L/R,0}  \label{baretrans}
\end{eqnarray}
we get
\begin{eqnarray}
\frac{\delta \Gamma}{\delta \phi^{-}_{0}}\, v_{0}
 + \sqrt{2}\,\biggl[
\frac{\delta \Gamma}{\delta {\bf d}_{L,0}} {\bf V}_{0}^{\dagger}{\bf
  u}_{L,0}
+{\bf \bar d}_{L,0}\V^{\dagger}_{0} \frac{\delta \Gamma}{\delta {\bf \bar u}_{L,0}}
\biggr] &=& 0 \nonumber \\
\frac{\delta \Gamma}{\delta \phi^{+}_{0}}\, v_{0}
 - \sqrt{2}\,\biggl[
\frac{\delta \Gamma}{\delta {\bf u}_{L,0}} {\bf V}_{0} {\bf d}_{L,0}
+{\bf \bar u}_{L,0}\V_{0} \frac{\delta \Gamma}{\delta {\bf \bar d}_{L,0}}
\biggr] &=& 0 \,.\label{wardmasseigen}
\end{eqnarray}
The bare biunitary transformation (\ref{baretrans}) 
relating bare mass and elektroweak 
eigenstates is defined to diagonalize the bare 
mass matrices. As a consequence in (\ref{wardmasseigen}) the 
bare CKM matrix
\begin{equation}
\V_{0} = \S_{u,0}^{L\,\dagger} \, \S_{d,0}^L 
\end{equation}
appears and the bare mass matrices of the mass eigenstates are diagonal.
In other words, in order not to violate the 
Ward identities mass renormalization has to be performed in such a way 
that no off
diagonal mass counterterms are needed.

\begin{figure}[t]
\begin{center}
\epsfig{bbllx=240,bblly=640,bburx=360,bbury=700,file=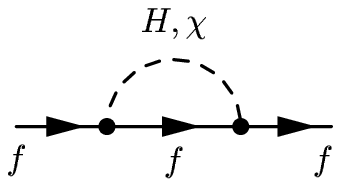}
\epsfig{bbllx=230,bblly=640,bburx=370,bbury=700,file=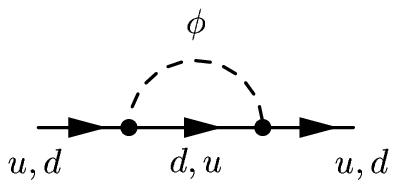}
\end{center}
\caption{\label{fig1} Quark self energy diagrams.}
\end{figure}

Upon functional differentiation with respect to 
up- and down quark field sources
the identities (\ref{wardmasseigen}) relate 
quark matrix elements of the charged current (\ref{L-int-geladen}) 
to the two point functions of the 
quarks. As a consequence the renormalization of the charge current
which defines the renormalization prescription for
 the CKM matrix is completely 
determined by the wave function renormalization constants of the left handed
quarks. This can be seen most easily as follows. 
If we assume that dimensional regularization respects the symmetry,
the Ward identities
are forminvariant under renormalization, i.e. they must hold also
for the renormalized fields and parameters
\begin{eqnarray}\label{wardmasseigenren}
\frac{\delta \Gamma}{\delta \phi^{-}}\, v
 + \sqrt{2}\,\biggl[
\frac{\delta \Gamma}{\delta {\bf d}_{L}} {\bf V}^{\dagger}{\bf
  u}_{L}
+{\bf \bar d}_{L}\V^{\dagger} \frac{\delta \Gamma}{\delta {\bf \bar u}_{L}}
\biggr] &=& 0 \nonumber \\
\frac{\delta \Gamma}{\delta \phi^{+}_{0}}\, v
 - \sqrt{2}\,\biggl[
\frac{\delta \Gamma}{\delta {\bf u}_{L}}{\bf V} {\bf d}_{L}
+{\bf \bar u}_{L}\V  \frac{\delta \Gamma}{\delta {\bf \bar d_{L}}}
\biggr] &=& 0 \,.
\end{eqnarray}
For (\ref{wardmasseigen},\ref{wardmasseigenren}) both to be valid 
the bare and the renormalized CKM matrix have to fulfil the relations
\begin{eqnarray} \label{ward1}
\V_0 &=& \sqrt{{\bf Z}_{u,L}}\,\, {\bf V} \,\, \sqrt{{\bf Z}_{d,L}}^{\,\,-1} \\
\sqrt{{\bf Z}_{u,L}^\dagger {\bf Z}_{u,L}}\, {\bf V} &=&  {\bf V} \,
\sqrt{{\bf Z}_{d,L}^\dagger {\bf Z}_{d,L}}   \label{ward2}
\end{eqnarray} 
where $\V_0$ is the bare, and $\V$ the renormalized CKM matrix and 
${\bf Z}_{u/d,L}$ are the matrices of wave function renormalization of 
the left handed
up and down quarks
\begin{eqnarray}
{\bf u}_{L,0} &=& \sqrt{{\bf Z}_{u,L}} \, {\bf u}_L \nonumber\\ 
{\bf d}_{L,0} &=& \sqrt{{\bf Z}_{d,L}} \, {\bf d}_L  \,.
\end{eqnarray}
Note that in perturbation theory we can evaluate the square root of 
these matrices
as well as we can invert them.
From equations (\ref{ward1}, \ref{ward2}) it follows that  
the unitarity of the bare CKM matrix implies the unitarity of the renormalized
CKM matrix (and {\it vice versa}) and has to 
be regarded
as a constraint on the ${\bf Z}_{u/d,L}$. 
The renormalization group equation for the CKM matrix 
\begin{equation}
\frac{d}{d\ln \mu}{\bf V} = {\bf \mbox{\boldmath{$\beta$}}}_{\bf V}
\end{equation}
is derived in the usual way by differentiating 
the bare CKM with respect to $\ln \mu$. Due to
(\ref{ward1}) this $\beta$-function can be expressed in terms of the
anomalous dimension matrices of the fields
\begin{equation}
{\bf \mbox{\boldmath{$\gamma$}}}_{u/d,L} = {\bf Z}_{u/d,L}^{-1}\frac{d}{d\ln \mu}{\bf Z}_{u/d,L}
\end{equation}
as
\begin{equation}
\mbox{\boldmath{$\beta$}}_{\bf V} = \frac{1}{2} [ {\bf V} {\bf \mbox{\boldmath{$\gamma$}}}_{d,L} - {\bf \mbox{\boldmath{$\gamma$}}}_{u,L} 
{\bf V} ] \,.
\end{equation}
The second of the Ward identities allows us to eliminate the hermitian
parts of the field anomalous dimensions and the final result for the 
$\beta$-function reads
\begin{equation}\label{betaV}
\mbox{\boldmath{$\beta$}}_{\bf V} = \frac{1}{4} [ {\bf V} ({\bf \mbox{\boldmath{$\gamma$}}}_{d,L} - {\bf \mbox{\boldmath{$\gamma$}}}_{d,L}^\dagger) 
          - ({\bf \mbox{\boldmath{$\gamma$}}}_{u,L} - {\bf \mbox{\boldmath{$\gamma$}}}_{u,L}^\dagger) {\bf V} ] \,.
\end{equation}
The appearence of only the antihermitian part is natural since upon
exponentiation, i.e. solving the renormalization group equation,
this yields the unitary contribution to the field renormalization matrix
which can be absorbed into a redefinition of the CKM matrix without destroying
its unitarity. This relation still holds to all orders and to evaluate it
at  one loop one needs to compute the divergent part of the quark 
self energies
shown in figure \ref{fig1}. 
From this we obtain for the hermitian and
antihermitian parts of the field anomalous dimensions for the left and right 
handed quarks
\begin{eqnarray}
 \left( \mbox{\boldmath{$\gamma$}}^{(1)}_{u,R} + \mbox{\boldmath{$\gamma$}}^{(1)\dagger}_{u,R} \right) &=& 
 \frac{1}{4\pi^2v^2}\,
2 \m_u^2 \\
 \left( {\bf \mbox{\boldmath{$\gamma$}}}^{(1)}_{u,L} + {\bf \mbox{\boldmath{$\gamma$}}}^{(1)\dagger}_{u,L} \right) &=&
 \frac{1}{4\pi^2v^2}\,
\bigg[ \m_u^2 + \V \m_d^2 \V^\dagger \bigg] \\
 \left( {\bf \mbox{\boldmath{$\gamma$}}}^{(1)}_{u,R} - {\bf \mbox{\boldmath{$\gamma$}}}^{(1)\dagger}_{u,R} \right)_{AC} &=& 
 \frac{1}{4\pi^2v^2}\,
6 \, \frac{m_{u,A} m_{u,C}}{m_{u,A}^2 - m_{u,C}^2}
\left( \V \m_d^2 \V^\dagger \right)_{AC} \quad A\not= C \\
 \left( {\bf \mbox{\boldmath{$\gamma$}}}^{(1)}_{u,L} - {\bf \mbox{\boldmath{$\gamma$}}}^{(1)\dagger}_{u,L} \right)_{AC} &=&
 \frac{1}{4\pi^2v^2}\,
3 \, \frac{m_{u,A}^2 + m_{u,C}^2}{m_{u,A}^2 - m_{u,C}^2}
\left( \V \m_d^2 \V^\dagger \right)_{AC} \quad A\not= C \\
 \left( {\bf \mbox{\boldmath{$\gamma$}}}^{(1)}_{u,R} - {\bf \mbox{\boldmath{$\gamma$}}}^{(1)\dagger}_{u,R} \right)_{AA} &=&
 \left( {\bf \mbox{\boldmath{$\gamma$}}}^{(1)}_{u,L} - {\bf \mbox{\boldmath{$\gamma$}}}^{(1)\dagger}_{u,L} \right)_{AA} =
 \frac{1}{4\pi^2v^2}\, 3i\,s_{u,A}\label{diago1} \\
 \left( {\bf \mbox{\boldmath{$\gamma$}}}^{(1)}_{d,R} + {\bf \mbox{\boldmath{$\gamma$}}}^{(1)\dagger}_{d,R} \right) &=& 
 \frac{1}{4\pi^2v^2}\,
2 \m_d^2\\ 
 \left( {\bf \mbox{\boldmath{$\gamma$}}}^{(1)}_{d,L} + {\bf \mbox{\boldmath{$\gamma$}}}^{(1)\dagger}_{d,L} \right) &=&
 \frac{1}{4\pi^2v^2}\,
\bigg[ \m_d^2 + \V^\dagger \m_u^2 \V \bigg] 
\end{eqnarray}
\begin{eqnarray}
 \left( {\bf \mbox{\boldmath{$\gamma$}}}^{(1)}_{d,R} - {\bf \mbox{\boldmath{$\gamma$}}}^{(1)\dagger}_{d,R} \right)_{BD} &=& 
 \frac{1}{4\pi^2v^2}\,
6 \, \frac{m_{d,B} m_{d,D}}{m_{d,B}^2 - m_{d,D}^2}
\left( \V^\dagger \m_u^2 \V \right)_{BD} \quad B\not= D \\
 \left( {\bf \mbox{\boldmath{$\gamma$}}}^{(1)}_{d,L} - {\bf \mbox{\boldmath{$\gamma$}}}^{(1)\dagger}_{d,L} \right)_{BD} &=&
 \frac{1}{4\pi^2v^2}\,
3 \, \frac{m_{d,B}^2 + m_{d,D}^2}{m_{d,B}^2 - m_{d,D}^2}
\left( \V^\dagger \m_u^2 \V \right)_{BD} \quad B\not= D \\
 \left( {\bf \mbox{\boldmath{$\gamma$}}}^{(1)}_{d,R} - {\bf \mbox{\boldmath{$\gamma$}}}^{(1)\dagger}_{d,R} \right)_{BB} &=&
 \left( {\bf \mbox{\boldmath{$\gamma$}}}^{(1)}_{d,L} - {\bf \mbox{\boldmath{$\gamma$}}}^{(1)\dagger}_{d,L} \right)_{BB} =
 \frac{1}{4\pi^2v^2}\, 3 i\,s_{d,B} \label{diago2}
\end{eqnarray}
where the capital letter indices run from $1$ to $3$, $m_{u/d,B}$ is
the mass of the up/down type quark of the B'th family and
the $s_{u/d,A}$ are explained after equation (\ref{sigmadecompose}).

From the self energy diagrams in figure \ref{fig1} we can also compute the
Higgs contribution to the mass renormalization. The bare mass matrices
are written as   
\begin{eqnarray}
\m_u^0 & = & \m_u + \dem_u \\
\m_d^0 & = & \m_d + \dem_d \, .
\end{eqnarray}
We choose the renormalization prescreption such that
the bare mass matrices and hence also $\dem_{u/d}$ are diagonal. 
This is possible, since we can absorb the off-diagonal elements
into the off-diagonal elements of
the antihermitian part of the right-handed wave function renormalization matrices
according to
\begin{eqnarray} 
&& \fmslash{p} \, \omega_+ \, \Si^{u,R}_{div}
+ \fmslash{p} \, \omega_- \, \Si^{u,L}_{div}
+ \m_u \, \omega_- \, \Si^{u,S}_{div}
+ \Si^{u,S}_{div} \, \m_u \, \omega_+ \nonumber \\
&& + \fmslash{p} \, \omega_+ \, \frac{1}{2} \left( \deZ^{u,R} +
\deZ^{u,R\,\dagger} \right)
+ \fmslash{p} \, \omega_- \, \frac{1}{2} \left( \deZ^{u,L} +
\deZ^{u,L\,\dagger} \right) \nonumber \\ &&
-\frac{1}{2} \left( \m_u \, \deZ^{u,R} + \deZ^{u,L\,\dagger} \, \m_u \right)
\omega_+
-\frac{1}{2} \left( \m_u \, \deZ^{u,L} + \deZ^{u,R\,\dagger} \, \m_u \right)
\omega_- \nonumber \\ &&
- \dem_u \, \omega_+ - \dem_u \, \omega_- = 0 \,.\label{bed-u-quarks}
\end{eqnarray}
The $\Si$ are given by a decomposition of the divergent parts
of the unrenormalized self energy diagrams in figure \ref{fig1}
\begin{equation}\label{sigmadecompose}
\Si^f_{div} = \fmslash{p} \, \omega_+ \, \Si^{f,R}_{div} +
\fmslash{p} \, \omega_- \, \Si^{f,L}_{div} +
\omega_+ \, \m_f \, \Si^{f,S}_{div} + \Si^{f,S}_{div} \, \m_f \, \omega_- \, . 
\end{equation}
The diagonal elements (\ref{diago1},\ref{diago2}), i.e. the parameters 
$s_{u/d,A}$, are not fixed.
This reflects the freedom to rephase the quark fields and 
from equation (\ref{bed-u-quarks}) it follows
that
the left and right handed contributions are identical but arbitrary.

\begin{figure}[t]
\begin{center}
\epsfig{bbllx=140,bblly=630,bburx=460,bbury=710,file=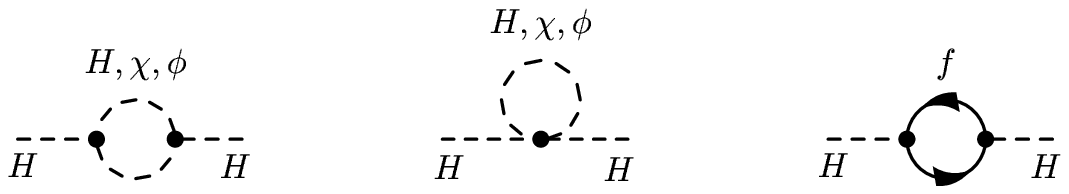}
\end{center}
\caption{\label{fig3}Self energy of the physical Higgs.}
\end{figure}

Using (\ref{betaV}) we derive the one loop
$\beta$-function for the CKM matrix as
\begin{eqnarray}
\left(\mbox{\boldmath{$\beta$}}^{(1)}_{\bf V}\right)_{AB} &=&
\frac{1}{16\pi^2v^2} \, 3 \,
\Bigg[ V_{AB} \, i \left(s_{d,B} - s_{u,A} \right) \nonumber \\
&& - \sum_{D\not=B} \sum_{C} V_{AD} V_{CD}^* V_{CB} \,
\frac{m_{d,B}^2 + m_{d,D}^2}{m_{d,B}^2 - m_{d,D}^2} \, m_{u,C}^2
\nonumber \\
&& - \sum_{C\not=A} \sum_{D} V_{AD} V_{CD}^* V_{CB} \,
\frac{m_{u,A}^2 + m_{u,C}^2}{m_{u,A}^2 - m_{u,C}^2} \, m_{d,D}^2 \Bigg]
\end{eqnarray} 
which matches exactly the full SM result \cite{denner-sack}.
The one loop contribution of the Higgs interactions 
to the mass renormalization using our prescription is
\begin{eqnarray}
\delta m_{u,A} &=&
\frac{1}{16\pi^2v^2}\,\Delta\,
\frac{3}{2} \, m_{u,A} \,
\bigg[ m_{u,A}^2 - \left( \V \m_d^2 \V^\dagger \right)_{AA} \bigg] \nonumber\\
\delta m_{d,B} &=&
\frac{1}{16\pi^2v^2}\,\Delta\,
\frac{3}{2} \, m_{d,B} \,
\bigg[ m_{d,B}^2 - \left( \V^\dagger \m_u^2 \V \right)_{BB} \bigg]
\end{eqnarray}
where $\Delta = 2/\varepsilon$.
Finally also the wave function renormalization constant $Z_H$ 
of the Higgs field is needed since this governs 
the renormalization of the vacuum expectation value
\begin{equation}
v_0 = Z_H \, v \,\,.
\end{equation}
From the Higgs self energy diagrams shown in figure \ref{fig3} 
we extract the one loop contribution to the 
Higgs field renormalization constant
\begin{equation}  \label{deltaZH} 
\delta Z_H = - \frac{1}{16\pi^2v^2}\,\Delta\,
2 N_c \, \tr \left(\m_u^2 + \m_d^2 \right) \,. 
\end{equation}

\begin{figure}[t]
\begin{center}
\epsfig{bbllx=220,bblly=630,bburx=370,bbury=710,file=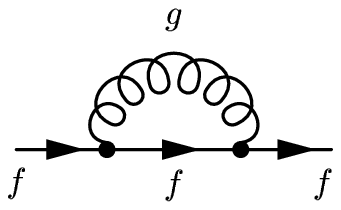}
\end{center}
\caption{\label{fig2} QCD contribution to the quark self energy.}
\end{figure}

As far as the elektroweak interaction is concerned this will also be the full
answer for $\mbox{\boldmath{$\beta$}}^{(1)}_{\bf V}$ 
since we are working at one loop. However, 
compared to the elektroweak contribution
a much larger
effect is the renormalization due to the strong interactions
which do not induce flavor mixing and thus modify only
the renormalization group functions for
the masses but not the $\beta$-function for the CKM matrix. 
Computing the QCD self energies shown in figure \ref{fig2} 
it turns out that we have 
\begin{eqnarray}\label{deltamf}
\delta m_{u,A} &=&
\frac{1}{16\pi^2v^2}\,\Delta\,
\frac{3}{2} \, m_{u,A} \,
\bigg[ m_{u,A}^2 - \left( \V \m_d^2 \V^\dagger \right)_{AA} \bigg]
- 2 \frac{\alpha_s}{\pi}\frac{\Delta}{2} \,m_{u,A}\nonumber \\
\delta m_{d,B} &=&
\frac{1}{16\pi^2v^2}\,\Delta\,
\frac{3}{2} \, m_{d,B} \,
\bigg[ m_{d,B}^2 - \left( \V^\dagger \m_u^2 \V \right)_{BB} \bigg]
- 2 \frac{\alpha_s}{\pi}\frac{\Delta}{2}\, m_{d,A} \,.
\end{eqnarray}

Thus we have gathered all the one loop contributions needed to perform 
a renormalization group study of the CKM matrix.

\section{Renormalization Group Flow}

We already derived the $\beta$-function for the CKM matrix and it only remains
to obtain the mass anomalous dimensions from (\ref{deltamf}) 
\begin{eqnarray}
\gamma_{u,A} &=& - m_{u,A}^{-1}\, \mu \frac{d}{d\mu} \, \delta m_{u,A}
\nonumber \\
\gamma_{d,B} &=& - m_{d,B}^{-1}\, \mu \frac{d}{d\mu} \, \delta m_{d,B} .
\end{eqnarray}
and the anomalous dimension $\gamma_v$ of the vacuum expectation value 
from (\ref{deltaZH})
\begin{equation} \label{def-gamma-v}
\gamma^{(1)}_v =\frac{\varepsilon}{2}  -
\mu \frac{d}{d\mu}  \delta Z_H 
\end{equation}
in order to
end up with a closed set of differential equations.
The term linear in $\varepsilon$ appears in (\ref{def-gamma-v}) 
since the vacuum expectation value changes its dimensionality
in dimensional regularization.
The total derivative with respect to $\ln \mu$ is in general 
\begin{eqnarray}
\mu\,\frac{d}{d\mu} &=& \mu\,\frac{\partial}{\partial\mu}
+ v \gamma_v \frac{\partial}{\partial v}
+ M_H^2 \gamma_H \frac{\partial}{\partial M_H^2}
+ M_\chi^2 \gamma_\chi \frac{\partial}{\partial M_\chi^2}
+ \m_u \mbox{\boldmath{$\gamma$}}_u \frac{\partial}{\partial \m_u} \nonumber \\
&&+ \m_d \mbox{\boldmath{$\gamma$}}_d \frac{\partial}{\partial \m_d} 
+ \mbox{\boldmath{$\beta$}}_\V \frac{\partial}{\partial \V} 
+ \beta_{\alpha_{s}}\frac{\partial}{\partial \alpha_{s}}
\end{eqnarray}
but to one loop this reduces to 
\begin{equation} \label{ableitung}
\mu\,\frac{d}{d\mu} 
= \frac{1}{\Delta} v \frac{\partial}{\partial v}
-2\alpha_{s} \frac{1}{\Delta} \frac{\partial}{\partial \alpha_{s}}\,.
\end{equation}
Since the various renormalization group functions are obtained by acting with the 
total derivative on one loop contributions 
to the bare parameters and renormalization constants, 
we have kept in (\ref{ableitung}) only the lowest order contributions, 
i.e. the terms linear in $\varepsilon$. These cancel with the divergent parts 
of the one loop terms in the bare quantities yielding a finite contribution to the 
one loop renormalization group functions.
In this way we obtain for the mass anomalous dimensions
\begin{eqnarray}
\gamma_{u,A}^{(1)} &=& \nonumber
\frac{1}{16\pi^2v^2}\, 3 \, 
\bigg[ m_{u,A}^2 - \left( \V \m_d^2 \V^\dagger \right)_{AA} \bigg]  
- 2 \frac{\alpha_s}{\pi} \\
\gamma_{d,B}^{(1)}  &=& \label{rgg-fkt-d}
\frac{1}{16\pi^2v^2}\, 3 \, 
\bigg[ m_{d,B}^2 - \left( \V^\dagger \m_u^2 \V \right)_{BB} \bigg] 
- 2 \frac{\alpha_s}{\pi}
\end{eqnarray}
and for the anomalous dimension of the vacuum expectation value
\begin{equation}
\gamma_v^{(1)} = - \frac{1}{16\pi^2v^2}\, \label{beta-fkt-v}
2 N_c \, \tr \left(\m_u^2 + \m_d^2 \right) 
\end{equation}
where we have dropped the term linear in $\varepsilon$. 

Thus the complete set of differential equations is 
\begin{eqnarray}
\mu \frac{d}{d\mu} \, v &=& \gamma^{(1)}_v \, v \label{runv}\\
\mu \frac{d}{d\mu} \, m_{u,A} &=& \gamma^{(1)}_{u,A}\, m_{u,A} \label{runmu} 
\end{eqnarray}
\begin{eqnarray}
\mu \frac{d}{d\mu} \, m_{d,B} &=& \gamma^{(1)}_{d,B}\, m_{d,B}\label{runmd} \\
\mu \frac{d}{d\mu} V_{AB} &=& \left(\mbox{\boldmath{$\beta$}}^{(1)}_{{\bf V}}\right)_{AB} \label{runV} \\ 
\mu \frac{d}{d\mu} \, \alpha_s &=& -2 \alpha_s \frac{\alpha_s}{\pi}\beta^{(1)}
\label{runalpha}
\end{eqnarray}
where $\beta^{(1)} = (33 - 2 n_f)/12$.
This set of equations is still valid for an arbitrary number of families.
The case of two families is practical trivial and hence we switch directly to 
the relevant case of three families.
Instead of working with the full matrices $V_{AB}$ we choose the standard
parametrization of the Particle Data Group \cite{pdg}
\begin{equation} \label{V-3-fam}
\V = \left(
\begin{array}{ccc}
\c_{12}\,\c_{13} & \s_{12}\,\c_{13} & \s_{13}\,e^{-i\delta_{13}} \\
-\s_{12}\,\c_{23}-\c_{12}\,\s_{23}\,\s_{13}\,e^{i\delta_{13}} &
\c_{12}\,\c_{23}-\s_{12}\,\s_{23}\,\s_{13}\,e^{i\delta_{13}} &
\s_{23}\,\c_{13} \\
\s_{12}\,\s_{23}-\c_{12}\,\c_{23}\,\s_{13}\,e^{i\delta_{13}} &  
-\c_{12}\,\s_{23}-\s_{12}\,\c_{23}\,\s_{13}\,e^{i\delta_{13}} &
\c_{23}\,\c_{13}
\end{array} \right)
\end{equation}
with 
\begin{eqnarray}
& \s_{12} = \sin (\theta_{12}), & \c_{12} = \cos (\theta_{12}),\nonumber \\
& \s_{13} = \sin (\theta_{13}), & \c_{13} = \cos (\theta_{13}),\nonumber \\
& \s_{23} = \sin (\theta_{23}), & \c_{23} = \cos (\theta_{23}) 
\end{eqnarray}
and write the renormalization group equations for the three angles $\theta_{ij}$
and the phase $\delta_{13}$
\begin{eqnarray}
\beta^{(1)}_{12} &=& \mu \, \frac{d}{d\mu} \, \theta_{12},\nonumber \\
\beta^{(1)}_{23} &=& \mu \, \frac{d}{d\mu} \, \theta_{23},\nonumber \\
\beta^{(1)}_{13} &=& \mu \, \frac{d}{d\mu} \, \theta_{13},\nonumber \\
\beta^{(1)}_\delta &=& \mu \, \frac{d}{d\mu} \, \delta_{13}.
\end{eqnarray}
The expressions for $\beta^{(1)}_{ij}$, $\beta^{(1)}_{\delta}$ and
$\gamma^{(1)}_{u/d,A}$ in terms of the angles $\theta_{ij}$ and the phase 
$\delta_{13}$ are quite lengthy and are deferred to the appendix.
Together with the equations for the masses (\ref{runmu}, \ref{runmd}), 
the vacuum expectation value (\ref{runv}) 
and the strong
coupling constant (\ref{runalpha}) 
this is a coupled system of $12$ differential equations 
which cannot be solved analytically
without approximations. We haved studied the exact solutions numerically and
found that they are reproduced with excellent accuracy  
by an approximative analytical solution to be discussed below. 

We shall study the renormalization group flow starting at the scale 
of top-quark mass; the discussion of the renormalization group 
flow below this scale is a separate issue since on then has to 
integrate out the top, bottom and charm quarks at the appropriate mass
scales. 

As initial values at the scale $\mu_{0} \approx m_t$ 
we choose \cite{pdg, ali-kayser}
\begin{equation}
  \begin{array}{rclp{2cm}rcl}
v &=& 245.3 \; \mathrm{GeV},& &\alpha_s &=& 0.109,\\
&&&&&&\\
\theta_{12} &=& 0.221,& &\theta_{23} &=& 0.039,\\
\theta_{13} &=& 0.0031,& &\delta_{13} &=& 1.26,\\
&&&&&&\\
m_{u,3} &\equiv& m_t = 165.8 \; \mathrm{GeV}.\\
  \end{array}
\end{equation}
The initial value of the running top mass $m_t(\mu)$ in the
$\overline{MS}$-scheme is related to the pole mass in the 
usual way
\begin{equation}
m_t(m_t^{pole}) = m_t^{pole}\,[1 - \frac{\alpha_s}{\pi}C_F]
\,,
\end{equation}
where $C_F = 4/3$ and 
the top quark pole mass is given by the experimental measured value
$m_t^{pole} = (173.8\pm 5.2)\,\mathrm{GeV}$. 

The masses of the light quarks ($u,c,d,s,b$) at $m_t$ have been evolved
from low scale by QCD corrections only:
\begin{equation}
  \begin{array}{rclp{2cm}rcl}
m_{u,1}&\equiv& m_u = 2.0 \;\mathrm{ MeV},&&
m_{d,1}&\equiv& m_d = 3.7 \;\mathrm{ MeV},\\
m_{u,2}&\equiv& m_c = 0.72 \;\mathrm{ GeV},&&
m_{d,2}&\equiv& m_s = 72 \;\mathrm{ MeV},\\
&&&& m_{d,3}&\equiv& m_b = 3.0 \;\mathrm{ GeV}.
  \end{array}
\end{equation}

Note that our results for the CKM matrix elements do not depend critically on
the exact values for the five light quark masses, hence we do not need to
include uncertainties in these masses.

The observed mass spectrum of the quarks together
with the hierarchy of the CKM angles allows us to construct an excellent
approximation for this system which can be solved analytically.

First of all we observe that the ratios 
$(m_{u/d,A}^2 + m_{u/d,C}^2)/(m_{u/d,A}^2 - m_{u/d,C})^2$ appearing in
$\beta_{ij}$ and $\beta_{\delta}$ 
are due to the large differences in the quark masses practically $\pm 1$.
Furthermore the renormalization group functions depend on $m_{u/d,A}^2/v^2$
which is extremely small except for the top quark. Hence we neglect
these terms and obtain
\begin{eqnarray}
\beta_{12} &=& \frac{3}{16\pi^2} \, \c_{12} \, \bigg[ \s_{12} \,
\left\{ \s_{23}^2 - \c_{23}^2 \, \s_{13}^2 \right\} - 2 \c_{12} \,
\c_{23} \, \s_{23} \, \s_{13} \, \cos (\delta_{13}) \bigg] \,
\frac{m_t^2}{v^2}\nonumber \\
\beta_{23} &=& \frac{3}{16\pi^2} \, \c_{23} \, \s_{23} \,
\frac{m_t^2}{v^2}\nonumber \\
\beta_{13} &=& \frac{3}{16\pi^2} \, \c_{23}^2 \, \c_{13} \, \s_{13} \,
\frac{m_t^2}{v^2}\nonumber \\
\beta_\delta &=& \frac{3}{16\pi^2} \, \frac{\c_{12} \, \c_{23} \,
\s_{23} \, \s_{13}}{\s_{12}} \, \sin (\delta_{13}) \,
\frac{m_t^2}{v^2}\nonumber \\
\gamma_v &=& - \frac{6}{16\pi^2} \, \frac{m_t^2}{v^2}\nonumber \\
\gamma_{u,1} &=& - 2 \frac{\alpha_s}{\pi}\nonumber \\
\gamma_{u,2} &=& - 2 \frac{\alpha_s}{\pi}\nonumber \\
\gamma_{u,3} &=& \frac{3}{16\pi^2} \, \frac{m_t^2}{v^2}
- 2 \frac{\alpha_s}{\pi} \\
\gamma_{d,1} &=& - \frac{3}{16\pi^2} \,
\bigg[ \s_{12}^2 \, \s_{23}^2 + \c_{12}^2 \, \c_{23}^2 \, \s_{13}^2
- 2\, \c_{12} \, \c_{23} \, \s_{12} \, \s_{23} \, \s_{13} \, \cos (\delta_{13})
\bigg] \, \frac{m_t^2}{v^2}\nonumber \\
&& - 2 \frac{\alpha_s}{\pi}\nonumber \\
\gamma_{d,2} &=& - \frac{3}{16\pi^2} \,
\bigg[ \c_{12}^2 \, \s_{23}^2 + \s_{12}^2 \, \c_{23}^2 \, \s_{13}^2
+ 2\, \c_{12} \, \c_{23} \, \s_{12} \, \s_{23} \, \s_{13} \, \cos (\delta_{13})
\bigg] \, \frac{m_t^2}{v^2}\nonumber \\
&& - 2 \frac{\alpha_s}{\pi}\nonumber \\
\gamma_{d,3} &=& - \frac{3}{16\pi^2} \, \c_{23}^2 \, \c_{13}^2 \,
\frac{m_t^2}{v^2} - 2 \frac{\alpha_s}{\pi}\nonumber \,.
\end{eqnarray}
Secondly we make use of the hierarchy of the CKM angles 
\begin{equation}
\theta_{12} = \mathcal{O}(10^{-1}), \quad
\theta_{23} = \mathcal{O}(10^{-2}), \quad
\theta_{13} = \mathcal{O}(10^{-3}) 
\end{equation}
thus keeping only terms of $\mathcal O(10^{-3})$ in $\gamma_{u/d,A}$,
$\beta_{ij}/\theta_{ij}$ and $\beta_{\delta}/\delta$.
From this we obtain the very simple system
\begin{eqnarray}
\beta_{23} &=& \frac{3}{16\pi^2} \, \frac{m_t^2}{v^2}\, \theta_{23}
 \nonumber \\
\beta_{13} &=& \frac{3}{16\pi^2} \, \frac{m_t^2}{v^2}\, \theta_{13}\nonumber 
\end{eqnarray}
\begin{eqnarray}
\gamma_{d,1} &=& - 2 \frac{\alpha_s}{\pi}\nonumber \\
\gamma_{d,2} &=& - 2 \frac{\alpha_s}{\pi}\nonumber \\
\gamma_{d,3} &=& - \frac{3}{16\pi^2} \, \frac{m_t^2}{v^2}
- 2 \frac{\alpha_s}{\pi} \label{naeh} \\
\gamma_{u,1} &=& - 2 \frac{\alpha_s}{\pi}\nonumber \\
\gamma_{u,2} &=& - 2 \frac{\alpha_s}{\pi}\nonumber \\
\gamma_{u,3} &=& \frac{3}{16\pi^2} \, \frac{m_t^2}{v^2}
- 2 \frac{\alpha_s}{\pi} \nonumber \\
\gamma_v &=& - 2 \, \frac{3}{16\pi^2} \, \frac{m_t^2}{v^2}\nonumber \,. 
\end{eqnarray}
The right hand side is determined by the top-Yukawa coupling
$Y_t = m_t/v$ for which we derive the
renormalization group equation 
\begin{equation}
\frac{d}{\ln \mu}\,Y_t = Y_t \left[ \frac{9}{16\pi^2}\, Y_t^2
- 2 \frac{\alpha_s(\mu)}{\pi}
\right]
\end{equation}
which can be solved analytically
\begin{eqnarray}
Y_t(\mu) &=& \Bigg[ Y_t^{-2}(\mu_0)
\left(\frac{\alpha_s(\mu_0)}{\alpha_s(\mu)}\right)^\frac{2}{\beta^{(1)}}
\nonumber \\
&&- \frac{9}{16\pi} \, \frac{1}{\beta^{(1)}-2} \, \frac{1}{\alpha_s(\mu)}
\left\{ 1-
\left( \frac{\alpha_s(\mu_0)}{\alpha_s(\mu)} \right)^{\frac{2}{\beta^{(1)}}-1}
\right\} \Bigg]^{-\frac{1}{2}}.
\end{eqnarray}
Where $\beta^{(1)}$ is the one-loop QCD $\beta$-function given after
equation (\ref{runalpha}). In leading logarithmic approximation the running of
the strong coupling constant
is given by the solution of (\ref{runalpha})
\begin{equation}
\alpha_s(\mu) = \frac{\alpha_s(\mu_0)}{1 +
\frac{2}{\pi}\,\beta^{(1)} \, \alpha_s(\mu_0) \ln \frac{\mu}{\mu_0}}\,.
\end{equation}
The differential equations (\ref{naeh}) for the masses, angles and the
vacuum expectation value can be written in the compact form
\begin{equation}\label{dgl2}
\frac{1}{y}\,\mu\,\frac{d}{d\mu}\,y =
\frac{3}{16\pi^2}\,c_y\,Y_t^2\left(\mu\right)
- q_y\,2 \frac{\alpha_s(\mu)}{\pi}
\end{equation}
where $y = \theta_{12}, \theta_{13}, \theta_{23}, \delta_{13},
m_u, m_d, m_c, m_s, m_b, m_t, v$ and 
\begin{equation} 
c_y = \left\{ \begin{array}{cl} 1 & \mbox{if }y= \theta_{23}, \theta_{13}, m_t
\\ -1 & \mbox{if } y=m_b \\ -2 & \mbox{if }y= v
\\ 0 & \mbox{if } y=\theta_{12}, \delta_{13}, m_u, m_d, m_s, m_c
\end{array} \right.
\end{equation}
and
\begin{equation}
q_y = \left\{ \begin{array}{cl} 0 & \mbox{if } y=\theta_{12}, \theta_{23},
\theta_{13}, \delta_{13}, v
\\ 1 & \mbox{if } y=m_u, m_d, m_s, m_c, m_b, m_t. \end{array} \right.
\end{equation}
The analytical solution of (\ref{dgl2}) reads
\begin{eqnarray}
y(\mu) &=& y(\mu_0) 
\Bigg[
1 + \frac{9}{16\pi} \,\frac{1}{\beta^{(1)}-2} \,\frac{1}{\alpha_s(\mu_0)}\,
\frac{m_t^2(\mu_0)}{v^2(\mu_0)} \left\{1 -
\left( \frac{\alpha_s(\mu_0)}{\alpha_s(\mu)} \right)^{1-\frac{2}{\beta^{(1)}}}
\right\} \Bigg]^{-\frac{1}{6}c_y}
\nonumber \\ && \hspace{2cm} {} \times \,
\Bigg[ \frac{\alpha_s(\mu_0)}{\alpha_s(\mu)}
\Bigg]^{-\frac{1}{\beta^{(1)}}q_y}
\end{eqnarray}
and in particular for the CKM matrix elements 
$V_{ub} \approx \theta_{13}e^{i\delta_{13}}$ and  
$V_{cb} \approx \theta_{23}$ 
\begin{eqnarray}
\frac{\left|V_{ub}(\mu)\right|}{\left|V_{ub}(\mu_0)\right|} &=&
\frac{\left|V_{cb}(\mu)\right|}{\left|V_{cb}(\mu_0)\right|}  \nonumber \\
&=& 
\Bigg[
1 + \frac{9}{16\pi} \,\frac{1}{\beta^{(1)}-2} \,\frac{1}{\alpha_s(\mu_0)}\,
\frac{m_t^2(\mu_0)}{v^2(\mu_0)} \left\{1 -
\left( \frac{\alpha_s(\mu_0)}{\alpha_s(\mu)} \right)^{1-\frac{2}{\beta^{(1)}}}
\right\} \Bigg]^{-\frac{1}{6}}\,. \nonumber \\ 
\end{eqnarray}

As a cross check we also solved the equations (\ref{runv}) to (\ref{runalpha})
numerically without any approximation. 
The deviation of the analytic solution from these numerical results are estimated by
the size of the integration interval of $\ln \mu$ of order $\mathcal{O}(10)$ times
the size of the terms of order $\mathcal{O}(10^{-4})$ neglected in the RG- and
$\beta$-functions. Indeed the deviation is less than a half percent for each
parameter.

The results of the analytic solution are shown in the figures \ref{fig4} to \ref{fig7}
in the next chapter together with the results of the full SM.

\section{Comparison with the full SM}

Up to now we have neglected the complete electroweak part and all leptons of the
SM. To be able to test the precission of our analytic approximation
we have solved the renormalization group equations numerically in the
full SM. This aapproch has also been chosen in
\cite{arason,olechowski,lindner1,lindner2,babu}.

Five additional parameters are entering the analysis. A convenient choice is the
coupling constant $g_Y$ of the $U(1)_Y$-hypercharge, the coupling constant $g_W$
of the $SU(2)_L$-weak interaction and the masses of the electron, the muon and the tau.
The one-loop results for the RG-functions of these parameters are
\begin{eqnarray}
\gamma_{g_Y^2} &=& \frac{g_Y^2}{4\pi^2} \, \frac{1}{12} \left(
1 + \sum_{fermions} (Y_L^2 + Y_R^2) \right) \\
\gamma_{g_W^2} &=& \frac{g_W^2}{4\pi^2} \,
\frac{-43 + 2 \, N_m}{12} \\
\gamma_l &=& \frac{g_Y^2}{4\pi^2} \, \frac{-11}{16} + \frac{g_W^2}{4\pi^2}
\, \frac{3}{16} +  \frac {3 m_l^2}{16\pi^2v^2} \quad (l=e,\mu,\tau)
\end{eqnarray}
where $N_m$ is the number of $SU(2)_L$ multiplets.
Furthermore, in the RG-functions of the quarks given in appendix an additional term appears
\begin{eqnarray}
\gamma_{u,A}^{elweak} &=& \frac{g_Y^2}{4\pi^2} \, \frac{-5}{48}
+ \frac{g_W^2}{4\pi^2}\, \frac{3}{16}\qquad (A = u,c,t)\\
\gamma_{d,B}^{elweak} &=& \frac{g_Y^2}{4\pi^2} \, \frac{7}{48}
+ \frac{g_W^2}{4\pi^2}\, \frac{3}{16}\qquad\, (B = d,s,b)
\end{eqnarray}
while for the vacuum expectation value the term
\begin{eqnarray}
\gamma_v^{elweak} &=& \frac{g_Y^2}{4\pi^2} \, \frac{1}{4}
+ \frac{g_W^2}{4\pi^2}\, \frac{3}{4}
\end{eqnarray}
must be added. As already pointed out in section 3 the $\beta$-functions for the CKM
parameters in the full SM are the same as in the ``ungauged'' model.
This follows from the fact that the electroweak contributions to the divergent part
of $\Sigma^R$ and $\Sigma^L$ are diagonal, flavor independent and real. In other words,
they do not have any antihermitian parts and thus cannot contribute to the
renormalization constant of the CKM matrix.

The RG-functions for the coupling constants decouple and lead to:
\begin{eqnarray}
\alpha_Y(\mu) = \frac{g_Y^2}{4\pi} &=& \frac{1}{\frac{1}{\alpha_Y(\mu_0)}
  - \frac{41}{12\pi} \ln(\frac{\mu}{\mu_0})} \qquad (\mu, \mu_0 \ge m_t)\\
\alpha_W(\mu) = \frac{g_W^2}{4\pi} &=& \frac{1}{\frac{1}{\alpha_W(\mu_0)}
  + \frac{19}{12\pi} \ln(\frac{\mu}{\mu_0})} \qquad (\mu, \mu_0 \ge m_t).
\end{eqnarray}

\begin{figure}[H] 
\begin{center}
\epsfig{bbllx=90,bblly=90,bburx=550,bbury=680,file=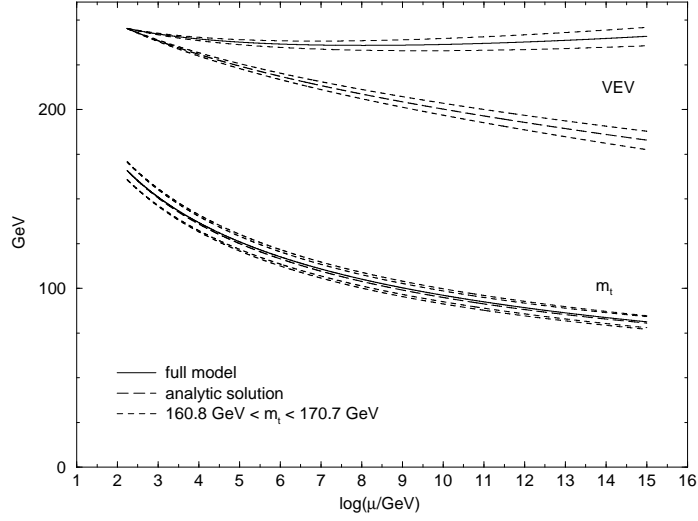,
        scale=0.435,angle=-90}
\end{center}
\caption{Renormalization group scaling of $m_t$ und $v$.
The width of the bands reflects the uncertainty of
$\pm \,5.2 \,\mathrm{GeV}$ in the top quark pole mass.}
\label{fig4}
\end{figure}

Figure \ref{fig4} shows the running of the top quark mass and the 
vacuum expectation value between $m_t$ and the large scale
$10^{15}\,\mathrm{GeV}$ according to our analytical approximation and the numerical
results in the full SM. While the vacuum expectation value differs
significantly, the top quark mass is practically the same in both cases.

\begin{figure}[H]
\begin{center}
\epsfig{bbllx=90,bblly=90,bburx=550,bbury=680,file=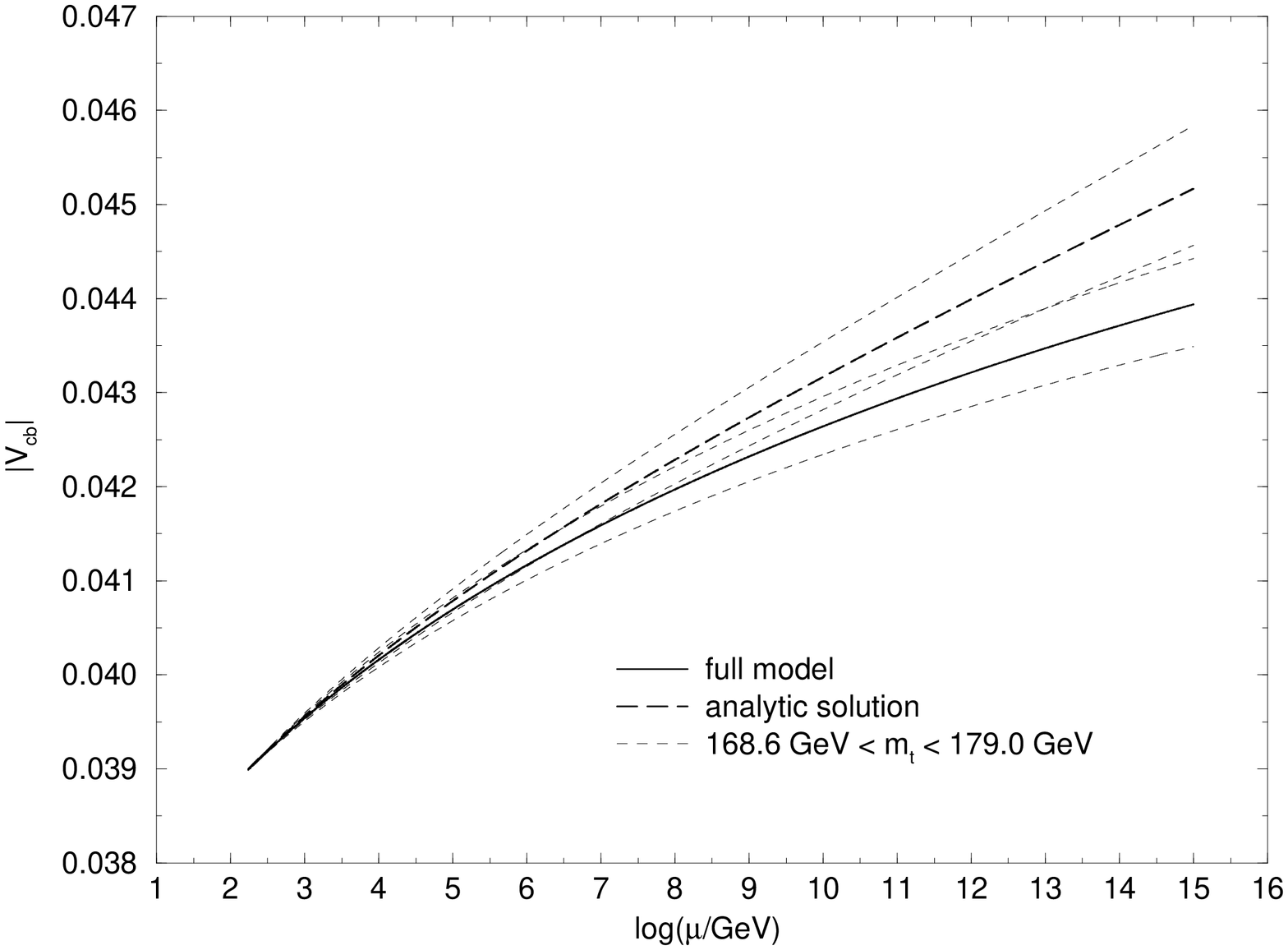,
        scale=0.435,angle=-90}
\end{center}
\caption{Renormalization group evolution of $V_{cb}$.}
\label{fig5}
\end{figure}

\begin{figure}[H]
\begin{center}
\epsfig{bbllx=90,bblly=90,bburx=550,bbury=680,file=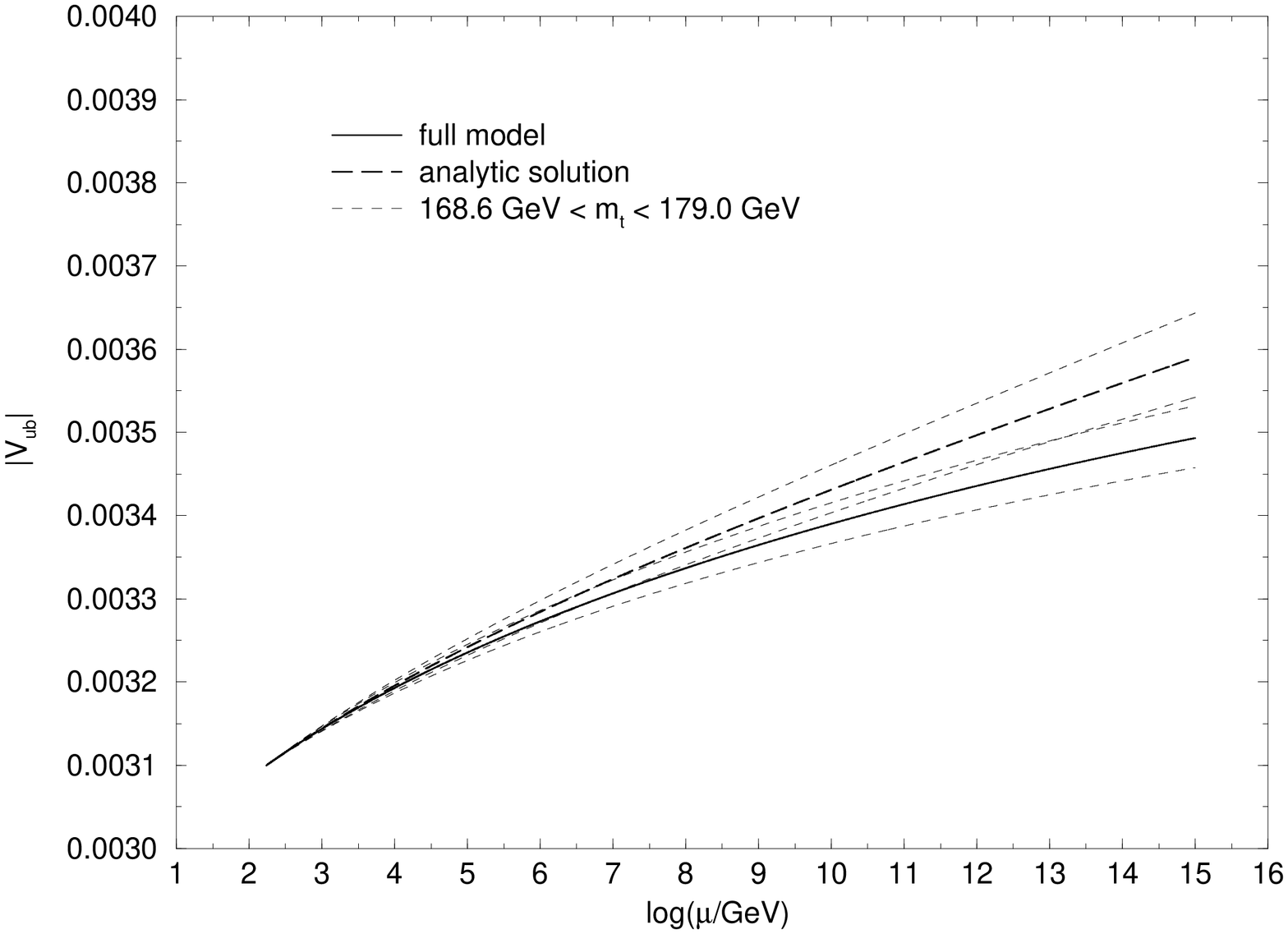,
        scale=0.435,angle=-90}
\end{center}
\caption{Renormalization group evolution of $V_{ub}$.}
\label{fig6}
\end{figure}

\begin{figure}[H]
\begin{center}
\epsfig{bbllx=90,bblly=90,bburx=550,bbury=680,file=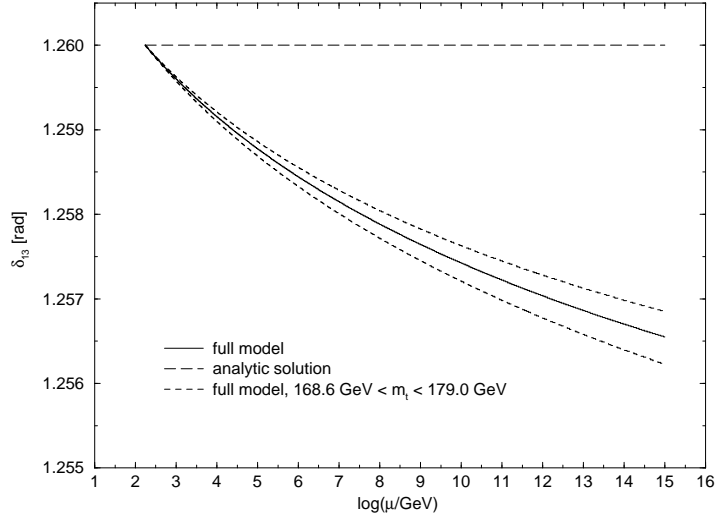,
        scale=0.435,angle=-90}
\end{center}
\caption{Renormalization group evolution of the phase $\delta_{13}$.}
\label{fig7}
\end{figure}

The renormalization group evolution of the CKM matrix elements $|V_{cb}|$ and $|V_{ub}|$
is plotted in figures \ref{fig5} and \ref{fig6}. Figure \ref{fig7} shows that the phase
$\delta_{13}$ is practically constant up to the GUT scale even in the full SM.

\begin{figure}[H]
\begin{center}
\epsfig{bbllx=90,bblly=90,bburx=550,bbury=680,file=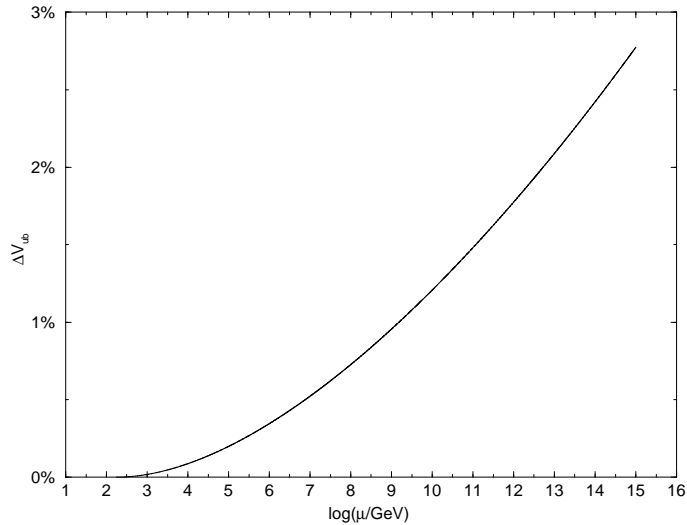,
        scale=0.435,angle=-90}
\end{center}
\caption{The relative difference between the analytic solution and the
numerical results in the full SM for $V_{ub}$.}
\label{fehler}
\end{figure}

Although the difference between the analytic solution and the full SM
is significant for the vacuum expectation value, the relative deviations for the
CKM matrix elements are less than $3\%$ for $|V_{cb}|$ and $|V_{ub}|$ in the whole range
of $m_t$ up to the GUT scale. In our approximation
these elements are identical with the parameters $\theta_{23}$ and $\theta_{13}$.
The approximation of constant parameters $\theta_{12}$ and $\delta_{13}$ is even better
with a relative difference of less than a half percent. As an example the relative
deviation of $|V_{ub}|$ is ploted in figure \ref{fehler}. The corresponding plot for
$V_{cb}$ is the same within the width of the lines.


\section{Conclusions}
We have studied the scale dependence of the CKM parameters
and quark masses. To one loop order it turned out that an ``ungauged''
SM
(i.e. only the Higgs) sector yields the same one loop result for
the $\beta$-function for the CKM matrix as the full
$SU(3)_C \otimes SU(2)_L \otimes U(1)_Y$ gauge theory.
This is obvious, since the renormalization of the Yukawa couplings due
to transverse gauge bosons is proportional to the unit matrix and hence
does not change $\delta \V$.
The renormalization of the CKM matrix is governed by a Ward identity
which allows us to express its renormalization solely in terms of
the quark self energies. This result is valid in the full SM and
we exploited it in the Higgs sector to derive the renormalization 
group functions
for the masses and CKM parameters.

Studying only the Higgs sector 
it becomes obvious that the running of the CKM matrix is
governed by the Yukawa couplings which are very small except for the
top quark. This motivates the limit in which all quark masses except the
one of the top are set to zero. However, it is well known that in such a limit
no mixing can occure due to the degeneracy of the down type quark masses.
In our results this is reflected by the appearence of the 
ratios $(m_{u/d,A}^2 + m_{u/d,C}^2)/(m_{u/d,A}^2 - m_{u/d,C}^2)$. The limiting
values of these single out a basis in flavor space relative to which
the CKM rotation can be defined. Using the physical values of the
masses this ratios are either $+1$ or $-1$ which simplifies the
renormalization group equations significantly. Putting in also the 
hierarchy of the CKM angles the renormalization group equations
can be solved analytically with an accuracy better than a few percent.

The renormalization group flow of the mixing between
the first two families turns out to be very small since the corresponding
Yukawa couplings are tiny.

For the third family the effects become sizeable and the parameters such
as $V_{cb}$ and $V_{ub}$ change at a level of $16\%$ between $m_t$
and the large scale $10^{15}\, \mathrm{GeV}$.
In the full SM this increase is reduced to $13\%$.
However, for the CKM matrix elements the analytic approximation solution differs
less than $3\%$ from the full numerical results.
Since in our approximation $V_{us}$ and the ratio $V_{ub}/V_{cb}$ do not
change with the renormalization scale all Wolfenstein parameters
\cite{wolfenstein} except $A=V_{cb}/V^2_{us}$ are scale independent.

\section*{Acknowledgements}
Ch.B. and T.M. gratefully acknowlegde the hospitality of the monestary
of Maria Laach and of Burg Liebenzell where parts of this work has been done.
Th.H. and B.P. acknowledge the support by the ``Studienstiftung des deutschen Volkes''.
The authors thank A. Denner and P. Gambino for interesting correspondence.
This work is supported by  
Graduiertenkolleg ``Elementarteilchenphysik an Beschleunigern'' and
DFG-Forschergruppe ``Quantenfeldtheorie, Computeralgebra und
Monte Carlo Simulation''.

\appendix
\section*{Appendix: RG- and $\beta$--functions in the particle data group
parametrization}
\begin{eqnarray} \label{beta_12}
\beta_{12} &=& \frac{3}{16\pi^2v^2}\, \Bigg\{
\frac{m_{d,1}^2 + m_{d,2}^2}{m_{d,1}^2 - m_{d,2}^2} \Bigg[ 
m_{u,1}^2 \, \s_{12} \, \c_{12} \, \c_{13}^2
\nonumber \\ && \mbox{}
+ m_{u,2}^2 \bigg( \Big\{  \s_{12}^2 - \c_{12}^2 \Big\} \, \s_{23} \,
\c_{23} \, \s_{13} \, \cos (\delta_{13})
+ \s_{12} \, \c_{12} \, \Big\{  \s_{23}^2 \, \s_{13}^2
- \c_{23}^2  \Big\} \bigg)
\nonumber \\ && \mbox{}
+ m_{u,3}^2 \bigg( -  \Big\{  \s_{12}^2
- \c_{12}^2 \Big\} \, \s_{23}
\, \c_{23} \, \s_{13} \, \cos (\delta_{13})
- \s_{12} \, \c_{12} \, \Big\{ \s_{23}^2 - \c_{23}^2 \,
\s_{13}^2 \Big\} \bigg) \Bigg] \nonumber \\
&& \mbox{} + \frac{m_{d,1}^2 + m_{d,3}^2}{m_{d,1}^2 - m_{d,3}^2} \Bigg[ 
m_{u,1}^2 \, \s_{12} \, \c_{12} \, \s_{13}^2
\nonumber \\
&&- m_{u,2}^2 \, \s_{12} \s_{23} \, \s_{13} \bigg( \c_{12} \, \s_{23} \, \s_{13}
+ \s_{12} \, \c_{23} \, \cos (\delta_{13}) \bigg)
\nonumber \\ && \mbox{}
- m_{u,3}^2 \, \s_{12} \c_{23} \, \s_{13} \bigg( \c_{12} \, \c_{23}
\, \s_{13} - \s_{12} \, \s_{23} \, \cos (\delta_{13}) \bigg) \Bigg]
\nonumber \\  
&& \mbox{} + \frac{m_{d,2}^2 + m_{d,3}^2}{m_{d,2}^2 - m_{d,3}^2} \Bigg[ 
- m_{u,1}^2 \, \s_{12} \, \c_{12} \, \s_{13}^2
\nonumber \\ && + m_{u,2}^2 \, \c_{12} \, \s_{23} \, \s_{13} \bigg( \s_{12} \, \s_{23} \,
\s_{13} - \c_{12} \, \c_{23} \, \cos (\delta_{13}) \bigg)
\nonumber \\ && \mbox{}
+ m_{u,3}^2 \, \c_{12} \, \c_{23} \, \s_{13} \bigg( \s_{12} \, \c_{23} \,
\s_{13} + \c_{12} \, \s_{23} \, \cos(\delta_{13}) \bigg) \Bigg]  \nonumber \\
&& \mbox{} + \frac{m_{u,1}^2 + m_{u,2}^2}{m_{u,1}^2 - m_{u,2}^2} \Bigg[
m_{d,1}^2 \, \c_{12} \, \c_{23} \bigg( \s_{12} \, \c_{23} + \c_{12} \,
\s_{23} \, \s_{13} \, \cos (\delta_{13}) \bigg)
\nonumber \\ && \mbox{}
- m_{d,2}^2 \, \s_{12} \, \c_{23}  \bigg( \c_{12} \, \c_{23} - \s_{12} \,
\s_{23} \, \s_{13} \, \cos (\delta_{13}) \bigg)
- m_{d,3}^2 \, \s_{23} \, \c_{23} \, \s_{13} \, \cos (\delta_{13}) \Bigg]
\nonumber \\
&& \mbox{} + \frac{m_{u,1}^2 + m_{u,3}^2}{m_{u,1}^2 - m_{u,3}^2} \Bigg[
m_{d,1}^2 \, \c_{12} \, \s_{23} \bigg( \s_{12} \, \s_{23} - \c_{12} \,
\c_{23} \, \s_{13} \, \cos( \delta_{13}) \bigg)
\nonumber \\ && \mbox{}
- m_{d,2}^2 \, \s_{12} \s_{23}  \bigg( \c_{12} \, \s_{23} + \s_{12} \,
\c_{23} \, \s_{13} \, \cos (\delta_{13}) \bigg)
+ m_{d,3}^2 \, \s_{23} \, \c_{23} \, \s_{13} \, \cos (\delta_{13}) \Bigg] 
\Bigg\}\nonumber\\
\end{eqnarray}
\begin{eqnarray}
\beta_{23} &=& \frac{3}{16\pi^2v^2}\, \Bigg\{
\frac{m_{d,1}^2 + m_{d,3}^2}{m_{d,1}^2 - m_{d,3}^2} \Bigg[
- m_{u,1}^2 \, \s_{12} \, \c_{12} \, \s_{13} \, \cos (\delta_{13})
\nonumber \\ && \mbox{}
+ m_{u,2}^2 \, \s_{12} \, \s_{23} \bigg( \s_{12} \, \c_{23} + \c_{12} \,
\s_{23} \, \s_{13} \, \cos (\delta_{13}) \bigg)
\nonumber \\ && \mbox{}
- m_{u,3}^2 \, \s_{12} \, \c_{23}  \bigg( \s_{12} \, \s_{23} - \c_{12} \,
\c_{23} \, \s_{13} \, \cos (\delta_{13}) \bigg) \Bigg] \nonumber \\
&& \mbox{} + \frac{m_{d,2}^2 + m_{d,3}^2}{m_{d,2}^2 - m_{d,3}^2} \Bigg[
m_{u,1}^2 \, \s_{12} \, \c_{12} \, \s_{13} \, \cos (\delta_{13})
\nonumber\\&&+ m_{u,2}^2 \, \c_{12} \, \s_{23}  \bigg( \c_{12} \, \c_{23} - \s_{12} \,
\s_{23} \, \s_{13} \, \cos (\delta_{13}) \bigg)
\nonumber \\ && \mbox{}
- m_{u,3}^2 \, \c_{12} \, \c_{23}  \bigg( \c_{12} \, \s_{23} + \s_{12} \,
\c_{23} \, \s_{13} \, \cos (\delta_{13}) \bigg) \Bigg] \nonumber \\
&& \mbox{} + \frac{m_{u,1}^2 + m_{u,2}^2}{m_{u,1}^2 - m_{u,2}^2} \Bigg[
- m_{d,1}^2 \, \c_{12} \, \c_{23} \, \s_{13} 
\bigg( \c_{12} \, \s_{23} \, \s_{13} + \s_{12} \, \c_{23} \, \cos (\delta_{13})
\bigg)
\nonumber \\ && \mbox{}
+ m_{d,2}^2 \, \s_{12} \, \s_{13} \bigg( \c_{12} \, \c_{23}^2 \,
\cos (\delta_{13}) - \s_{12} \, \s_{23} \, \c_{23} \, \s_{13} \, \bigg)
+ m_{d,3}^2 \, \s_{23} \, \c_{23} \, \s_{13}^2 \Bigg] \nonumber \\ 
&& \mbox{} + \frac{m_{u,1}^2 + m_{u,3}^2}{m_{u,1}^2 - m_{u,3}^2} \Bigg[
m_{d,1}^2 \, \c_{12} \, \s_{23} \, \s_{13} \bigg( \c_{12} \, \c_{23} \, \s_{13}
- \s_{12} \, \s_{23} \, \cos(\delta_{13}) \bigg)
\nonumber \\ && \mbox{}
+ m_{d,2}^2 \, \s_{12} \, \s_{23} \, \s_{13} \bigg( \s_{12} \, \c_{23} \,
\s_{13} + \c_{12} \, \s_{23} \, \cos (\delta_{13}) \bigg)
- m_{d,3}^2 \, \s_{23} \, \c_{23} \, \s_{13}^2 \Bigg] \nonumber \\
&& \mbox{} + \frac{m_{u,2}^2 + m_{u,3}^2}{m_{u,2}^2 - m_{u,3}^2} \Bigg[
m_{d,1}^2 \bigg( \s_{23} \, \c_{23} \, \Big\{ \s_{12}^2 - \c_{12}^2 \,
\s_{13}^2 \Big\} \nonumber \\&&+ \s_{12} \, \c_{12} \, \s_{13} \,
\Big\{ \s_{23}^2 - \c_{23}^2 \Big\} \, \cos (\delta_{13}) \bigg)
\nonumber \\ && \mbox{}
+ m_{d,2}^2 \bigg( \s_{23} \, \c_{23} \, \Big\{ \c_{12}^2 - \s_{12}^2 \,
\s_{13}^2 \, \Big\} - \s_{12} \, \c_{12} \, \s_{13} \,
\Big\{ \s_{23}^2 - \c_{23}^2 \Big\} \, \cos (\delta_{13}) \bigg)
\nonumber \\ && \mbox{} 
- m_{d,3}^2 \, \s_{23} \, \c_{23} \, \c_{13}^2 \Bigg] \Bigg\}
\end{eqnarray}
\begin{eqnarray}
\beta_{13} &=& \frac{3}{16\pi^2v^2}\, \Bigg\{
\frac{m_{d,1}^2 + m_{d,3}^2}{m_{d,1}^2 - m_{d,3}^2} \Bigg[
m_{u,1}^2 \, \c_{12}^2 \, \s_{13} \, \c_{13}
\nonumber \\&&- m_{u,2}^2 \, \c_{12} \s_{23} \, \c_{13} \bigg( \c_{12} \, \s_{23} \,
\s_{13} + \s_{12} \, \c_{23} \, \cos (\delta_{13}) \bigg) \
\nonumber \\ && \mbox{}
- m_{u,3}^2 \, \c_{12} \, \c_{23} \, \c_{13} \bigg( \c_{12} \, \c_{23} \,
\s_{13} - \s_{12} \, \s_{23} \, \cos (\delta_{13}) \bigg) \Bigg] \nonumber \\
&& \mbox{} + \frac{m_{d,2}^2 + m_{d,3}^2}{m_{d,2}^2 - m_{d,3}^2} \Bigg[
m_{u,1}^2 \, \s_{12}^2 \, \s_{13} \, \c_{13}
\nonumber\\&&- m_{u,2}^2 \, \s_{12} \, \s_{23} \, \c_{13} \bigg( \s_{12} \, \s_{23} \,
\s_{13} - \c_{12} \, \c_{23} \, \cos(\delta_{13}) \bigg)
\nonumber \\ && \mbox{}
- m_{u,3}^2 \, \s_{12} \, \c_{23} \, \c_{13} \bigg( \s_{12} \, \c_{23} \,
\s_{13} + \c_{12} \, \s_{23} \, \cos (\delta_{13}) \bigg)  \Bigg] \nonumber \\
&& \mbox{} + \frac{m_{u,1}^2 + m_{u,2}^2}{m_{u,1}^2 - m_{u,2}^2} \Bigg[
m_{d,1}^2 \, \c_{12} \, \s_{23} \, \c_{13} \bigg( \c_{12} \, \s_{23} \, \s_{13}
+ \s_{12} \, \c_{23} \, \cos (\delta_{13}) \bigg)
\nonumber \\ && \mbox{}
+ m_{d,2}^2 \, \s_{12} \, \s_{23} \, \c_{13} \bigg( \s_{12} \, \s_{23} \,
\s_{13} - \c_{12} \, \c_{23} \, \cos(\delta_{13}) \bigg)
- m_{d,3}^2 \, \s_{23}^2 \, \s_{13} \, \c_{13}  \Bigg] \nonumber \\
&& \mbox{} + \frac{m_{u,1}^2 + m_{u,3}^2}{m_{u,1}^2 - m_{u,3}^2} \Bigg[
m_{d,1}^2 \, \c_{12} \, \c_{23} \, \c_{13} \bigg( \c_{12} \, \c_{23} \, \s_{13}
- \s_{12} \, \s_{23} \, \cos (\delta_{13}) \bigg)
\nonumber \\ && \mbox{}
+ m_{d,2}^2 \, \s_{12} \, \c_{23} \, \c_{13} \bigg( \s_{12} \, \c_{23} \,
\s_{13} + \c_{12} \, \s_{23} \, \cos (\delta_{13}) \bigg)
- m_{d,3}^2 \, \c_{23}^2 \, \s_{13} \, \c_{13} \Bigg] \Bigg\}\nonumber\\
\end{eqnarray}
\newpage
\begin{eqnarray}
  \thispagestyle{empty}
 \beta_\delta &=& \frac{3}{16\pi^2v^2} \, \sin (\delta_{13}) \, \Bigg\{
\frac{m_{d,1}^2 + m_{d,2}^2}{m_{d,1}^2 - m_{d,2}^2} \, 
\Bigg[
m_{u,2}^2 \, \frac{ \s_{23} \, \c_{23} \, \s_{13}}{\s_{12} \, \c_{12}} 
- m_{u,3}^2 \, \frac{ \s_{23} \, \c_{23} \, \s_{13} }{ \s_{12} \, \c_{12} }
\Bigg] \nonumber \\
&& \mbox{} +  \frac{m_{d,1}^2 + m_{d,3}^2}{m_{d,1}^2 - m_{d,3}^2} \, \Bigg[ 
m_{u,1}^2 \, \frac{ \s_{12} \, \c_{12} \, \s_{13} \, \Big\{ \c_{23}^2 
- \s_{23}^2 \Big\} }{ \s_{23} \, \c_{23} }
\nonumber \\ && \mbox{}
+ m_{u,2}^2 \, \s_{12} \, \s_{23} \frac{ \Big\{ \c_{12}^2 - \c_{23}^2 \Big\} \,
\s_{13}^2 + \c_{12}^2 \, \c_{23}^2 \, \c_{13}^2 }{ \c_{12} \, \c_{23} \,
\s_{13}}
\nonumber\\&&+ m_{u,3}^2 \, \s_{12} \, \c_{23} \frac{ \s_{23}^2 \, \s_{13}^2
+ \c_{12}^2 \, \Big\{ \c_{23}^2 \, \c_{13}^2
- 1 \Big\}}{ \c_{12} \, \s_{23} \, \s_{13}}
\Bigg] \nonumber \\
&& \mbox{} + \frac{m_{d,2}^2 + m_{d,3}^2}{m_{d,2}^2 - m_{d,3}^2} \, \Bigg[
m_{u,1}^2\, \frac{ \s_{12} \, \c_{12} \, \s_{13} \, \Big\{  \c_{23}^2
- \s_{23}^2 \Big\} }{ \s_{23} \, \c_{23}}
\nonumber \\ && \mbox{}
+ m_{u,2}^2 \, \c_{12} \, \s_{23} \frac{ \Big\{ \c_{12}^2 - \c_{13}^2
\Big\} \, \c_{23}^2 - \s_{12}^2 \, \s_{23}^2 \, \s_{13}^2 }{ \s_{12} \,
\c_{23} \, \s_{13}}
\nonumber \\ && \mbox{}
+ m_{u,3}^2 \, \c_{12} \, \c_{23}
\frac{ \c_{23}^2 \, \s_{13}^2 + \s_{23}^2 \, \c_{13}^2 + \c_{12}^2 \,
\Big\{ \c_{23}^2 \, \c_{13}^2 - 1 \Big\} }{ \s_{12} \, \s_{23} \, \s_{13} }
\Bigg] \nonumber \\
&& \mbox{} + \frac{m_{u,1}^2 + m_{u,2}^2}{m_{u,1}^2 - m_{u,2}^2} \, \Bigg[
- m_{d,1}^2 \, \c_{12} \, \c_{23}
\frac{ \c_{12}^2 \, \s_{13}^2 + \s_{12}^2 \, \c_{13}^2 + \c_{23}^2 \,
\Big\{ \c_{12}^2 \, \c_{13}^2 - 1 \Big\} }{ \s_{12} \, \s_{23} \, \s_{13}}
\nonumber \\ && \mbox{}
-  m_{d,2}^2 \, \s_{12} \, \c_{23} \frac{ \c_{12}^2 \, \Big\{
\c_{23}^2 - \c_{13}^2 \Big\} - \s_{12}^2 \, \s_{23}^2 \, \s_{13}^2 }{
\c_{12} \, \s_{23} \, \s_{13}}
\nonumber\\&&+  m_{d,3}^2 \, \frac{ \s_{23} \, \c_{23} \, \s_{13} \, \Big\{  \c_{12}^2
- \s_{12}^2 \Big\}}{ \s_{12} \, \c_{12}} \Bigg] \nonumber \\
&& \mbox{} + \frac{m{_u1}^2 + m_{u,3}^2}{m{_u1}^2 - m_{u,3}^2} \, \Bigg[
- m_{d,1}^2 \, \c_{12} \, \s_{23}
\frac{ \s_{12}^2 \, \s_{13}^2 + \c_{23}^2 \, \Big\{ \c_{12}^2 \, \c_{13}^2
- 1 \Big\} }{ \s_{12} \, \c_{23} \, \s_{13} }
\nonumber \\ && \mbox{}
- m_{d,2}^2 \, \s_{12} \, \s_{23} \frac{ - \Big\{ \c_{12}^2  - \c_{23}^2 \Big\}
\, \s_{13}^2 + \c_{12}^2 \, \c_{23}^2 \, \c_{13}^2 }{ \c_{12} \, \c_{23} \,
\s_{13} }
\nonumber\\&&- m_{d,3}^2 \, \frac{ \s_{23} \, \c_{23} \, \s_{13} \, \Big\{ \c_{12}^2
- \s_{12}^2 \Big\} }{ \s_{12} \, \c_{12}} \Bigg] \nonumber \\
&& \mbox{} + \frac{m_{u,2}^2 + m_{u,3}^2}{m_{u,2}^2 - m_{u,3}^2} \, \Bigg[
m_{d,1}^2 \, \frac{ \s_{12} \, \c_{12} \, \s_{13} }{ \s_{23} \, \c_{23} } 
- m_{d,2}^2 \, \frac{ \s_{12} \, \c_{12} \, \s_{13} }{ \s_{23} \, \c_{23}}
\Bigg] \Bigg\}.
\end{eqnarray}
\newpage
\begin{eqnarray}
\thispagestyle{empty}
  \gamma_v &=&
- \frac{2 N_c}{16\pi^2v^2} \bigg[ m_{u,1}^2 + m_{u,2}^2 + m_{u,3}^2 +
m_{d,1}^2 + m_{d,2}^2 + m_{d,3}^2 \bigg]
\\ \nonumber && \\
\gamma_{u,1} &=&
\frac{3}{16\pi^2v^2} \Bigg[ m_{u,1}^2 - \c_{12}^2 \, \c_{13}^2 \, m_{d,1}^2
- \s_{12}^2 \, \c_{13}^2 \, m_{d,2}^2 - \s_{13}^2 \, m_{d,3}^2 \bigg] 
- 2 \frac{\alpha_s}{\pi} 
\\ \nonumber && \\
\gamma_{u,2} &=&
\frac{3}{16\pi^2v^2} \Bigg[ m_{u,2}^2
- \bigg( \s_{12}^2 \, \c_{23}^2 + \c_{12}^2 \, \s_{23}^2 \, \s_{13}^2
+ 2 \, \s_{12} \, \c_{12} \, \s_{23} \, \c_{23} \, \s_{13} \,
\cos (\delta_{13}) \bigg) \, m_{d,1}^2
\nonumber \\ && \mbox{}
- \bigg( \c_{12}^2 \, \c_{23}^2 + \s_{12}^2 \, \s_{23}^2 \, \s_{13}^2
- 2 \, \s_{12} \, \c_{12} \, \s_{23} \, \c_{23} \, \s_{13} \,
\cos (\delta_{13}) \bigg) m_{d,2}^2
\nonumber\\&&- \s_{23}^2 \, \c_{13}^2 \, m_{d,3}^2 \Bigg]
- 2 \frac{\alpha_s}{\pi} 
\\ \nonumber && \\
\gamma_{u,3} &=&
\frac{3}{16\pi^2v^2} \Bigg[ m_{u,3}^2
- \bigg( \s_{12}^2 \, \s_{23}^2 + \c_{12}^2 \, \c_{23}^2 \, \s_{13}
- 2 \, \s_{12} \, \c_{12} \, \s_{23} \, \c_{23} \, \s_{13} \,
\cos (\delta_{13}) \bigg) \, m_{d,1}^2
\nonumber \\ && \mbox{}
- \bigg( \c_{12}^2 \, \s_{23}^2 + \s_{12}^2 \, \c_{23}^2 \, \s_{13}^2 
+ 2 \, \s_{12} \, \c_{12} \, \s_{23} \, \c_{23} \, \s_{13} \,
\cos (\delta_{13}) \bigg) \, m_{d,2}^2
\nonumber\\&&- \c_{23}^2 \, \c_{13}^2 \, m_{d,3}^2 \Bigg]
- 2 \frac{\alpha_s}{\pi} 
\\ \nonumber && \\
\gamma_{d,1} &=&
\frac{3}{16\pi^2v^2} \Bigg[ m_{d,1}^2 - \c_{12}^2 \, \c_{13}^2 \, m_{u,1}^2
\nonumber \\ && \mbox{} 
- \bigg( \s_{12}^2 \, \c_{23}^2 + \c_{12}^2 \, \s_{23}^2 \, \s_{13}^2
+ 2 \, \s_{12} \, \c_{12} \, \s_{23} \, \c_{23} \, \s_{13} \,
\cos (\delta_{13}) \bigg) \, m_{u,2}^2
\nonumber \\ && \mbox{}
- \bigg( \s_{12}^2 \, \s_{23}^2 + \c_{12}^2 \, \c_{23}^2 \, \s_{13}^2
- 2 \, \s_{12} \, \c_{12} \, \s_{23} \, \c_{23} \, \s_{13} \,
\cos (\delta_{13}) \bigg) \, m_{u,3}^2 \Bigg]
- 2 \frac{\alpha_s}{\pi}
\\ \nonumber && \\
\gamma_{d,2} &=&
\frac{3}{16\pi^2v^2} \Bigg[ m_{d,2}^2 - \s_{12}^2 \, \c_{13}^2 \, m_{u,1}^2
\nonumber \\ && \mbox{} 
- \bigg( \c_{12}^2 \, \c_{23}^2 + \s_{12}^2 \, \s_{23}^2 \, \s_{13}^2
- 2 \, \s_{12} \, \c_{12} \, \s_{23} \, \c_{23} \, \s_{13} \,
\cos (\delta_{13}) \bigg) \, m_{u,2}^2
\nonumber \\ && \mbox{}
- \bigg( \c_{12}^2 \, \s_{23}^2 + \s_{12}^2 \, \c_{23}^2 \, \s_{13}^2
+ 2 \, \s_{12} \, \c_{12} \, \s_{23} \, \c_{23} \, \s_{13} \,
\cos (\delta_{13}) \bigg) \, m_{u,3}^2 \Bigg]
- 2 \frac{\alpha_s}{\pi}
\\ \nonumber && \\
\gamma_{d,3} &=&
\frac{3}{16\pi^2v^2} \Bigg[ m_{d,3}^2 - \s_{13}^2 \, m_{u,1}^2
- \s_{23}^2 \, \c_{13}^2 \, m_{u,2}^2 
- \c_{23}^2 \, \c_{13}^2 \, m_{u,3}^2 \Bigg]
- 2 \frac{\alpha_s}{\pi}
\end{eqnarray}

\end{document}